\documentclass[floatfix, twocolumn,amsmath,amssymb, aps,prb,superscriptaddress]{revtex4-2}
\usepackage{mathrsfs}
\usepackage{amsmath}
\usepackage{ragged2e}

\usepackage{caption}
\usepackage{subcaption}
\usepackage{graphicx}
\usepackage{hyperref}
\usepackage{stfloats}
\usepackage{ulem}
\hypersetup{ colorlinks, citecolor=blue, linkcolor=blue, urlcolor=blue}

\DeclareCaptionLabelFormat{nocaption}{}

\begin{document}
	\title{Novel topological phases of a semi-Dirac Chern insulator in presence of extended range hopping}
	
	\author{Sayan Mondal}
	\affiliation{Department of Physics, Indian Institute of Technology Guwahati, 
		Guwahati 781039, Assam, India}
	\author{Saurabh Basu}
	\affiliation{Department of Physics, Indian Institute of Technology Guwahati, 
		Guwahati 781039, Assam, India}

	\begin{abstract}
	We study topological properties and the topological phase transitions therein for a semi-Dirac Haldane model on a honeycomb lattice in presence of an extended range (third neighbour) hopping. While in the absence of a third neighbour hopping, $t_3$, the system exhibits gapless electronic spectrum, its presence creates an energy gap in the dispersion. However, the nature of the spectral gap, that is, whether it is trivial or topological needs to be ascertained. We find that the answer depends on the value of $t_3$, and its interplay with the value of the onsite potential that breaks the sublattice symmetry, namely, Semenoff mass ($\Delta$). To elucidate our findings on the topological phases, we demonstrate two kinds of phase diagrams using the available parameter space, one in which the phases are shown in the $\Delta$-$t_3$ plane, and the other one in a more familiar $\Delta$-$\phi$ plane ($\phi$ being the Haldane flux). The phase diagrams depict presence of Chern insulating lobes comprising of Chern numbers $\pm2$ and $\pm1$ for certain values of $t_3$, along with trivial insulating regions (zero Chern number). Thus there are phase transitions from one topological regime to another which are characterized by abrupt changes in the values of the Chern number. To support the existence of the topological phases, we compute the counter propagating chiral edge modes in a ribbon geometry. Finally, the anomalous Hall conductivity shows plateaus either at $e^2/h$ or $2e^2/h$ corresponding to these topological phases. 
		
	\end{abstract}
	\maketitle
	
	\section{Introduction}
	In condensed matter systems, the information whether a material possesses a topological phase, has been an immense interest since the discovery of quantum Hall effect (QHE) \cite{klitzing}. The latter demonstrates that in presence of a strong magnetic field, the Hall conductivity of a two-dimensional electron gas acquires a series of plateaus quantized in unit of $e^2/h$. This quantization was due to the presence of the discrete magnetic Bloch bands \cite{thouless1982,thouless1983,avron1983,kohomoto1985,Niu1985,prange1990} or the Landau levels \cite{laughlin,ilani2004,trugman1983,tong,vasil1985} owing to the presence of a magnetic flux. The topological invariant that defines the quantization of the Hall conductivity, from a general perspective, is known as the Thouless-Kohmoto-Nightingale-Nijs (TKNN) invariant \cite{thouless1982}. 
		
	An external  magnetic field, initially appeared to be necessary to achieve QHE, however, Haldane had proposed that even in absence of an external magnetic flux, QHE still can be observed \cite{Haldane1988}. He introduced a direction dependent complex next nearest neighbour hopping in a honeycomb lattice, such as graphene, which breaks the time reversal symmetry (TRS). This broken TRS is the only necessary criterion to observe QHE. The model proposed by Haldane is a two band system, with the bands being characterized by a topological invariant known as the Chern number, and these quantized (integer) values of the Chern number yield a plateau in the Hall conductivity when the Fermi energy lies in the bulk gap. Further, a non-zero value of the Chern number and hence the quantized value of the Hall conductivity can be seen for a finite value of the phase, $\phi$ (we shall call it Haldane flux) of the complex next nearest neighbour hopping. A non-zero value of the Semenoff mass, $\Delta$, that breaks the sublattice symmetry in graphene, opens or closes a gap in the bandstructure \cite{semenhoff}. The variation of the Semenoff mass with the Haldane flux presents a phase diagram that encodes opening and closing of the band gap alternately at the two Dirac points, which are usually called as the $\mathbf{K}$ and the $\mathbf{K}^\prime$ points \cite{Haldane1988,vanderbilt2006}.
	
	In recent years, exploration of the topological properties associated with the Haldane model have progressed rapidly in quantum many body systems \cite{hasan2010,ando2013,qi2011,moore2010} both from the experimental and the theoretical perspectives. The study in the two dimensional Dirac systems have also been explored, such as, Fe based ferromagnetic insulators, $\mathrm{XFe_2(PO_4)_2}$, where X may be Cs, K La etc. \cite{kim2017}, in Dice lattice \cite{Kapri2020} etc. which host isotropic low energy Dirac like dispersions. However, there exists certain other materials that display anisotropic dispersions at low energies, for example, say quadratic along one direction in the Brillouin zone (BZ), and linear along the other one \cite{Dietl2008,Banerjee2009,zieglar2017}, which are known as the semi-Dirac systems. The semi-Dirac dispersions have been found in a variety of materials, such as, phosphorene under pressure and doping \cite{castro_2014,guan_2014}, electric fields \cite{katnelson_2015, katnelson_2016}, in multilayered structures of $\mathrm{TiO}_2/\mathrm{VO}_2$ \cite{pickett2009,pickett2010}, $\mathrm{BEDT}$-$\mathrm{TTF_2I_3}$ organic salts under pressure \cite{Suzumura2013,hasegawa2006}, oxidized silicene layer \cite{zhang_2017}, deformed graphene \cite{montambaux_2009} etc. Experimentally, the semi-Dirac dispersion has been observed in layers of black phosphorene obtained via \textit{in situ} doping of potassium atoms \cite{kim2015}. 
	
	A natural question arises: whether the semi-Dirac system will show similar topological phases as their Dirac counter part. To have an insight on the answer to this, we wish to explore whether and how the lowering of symmetry induced by anisotropic dispersion modifies the topological properties of the system. However, to achieve a topological phase, we need to break the TRS of the system, either by including a perpendicular magnetic field or via adding the complex second neighbour hopping (the Haldane term). Here, we choose the latter option. However, unlike the Dirac case, addition of the complex second neighbour hopping in the semi-Dirac system does not open a gap in the electronic spectrum, and hence the system still remains a semi-metal, with the conduction and the valence bands touching each other at a point intermediate to the $\mathbf{K}$ and the $\mathbf{K}^\prime$ points in the BZ (the $\mathbf{M}$ point). A little introspection reveals that, in such a scenario we can add a real third neighbour hopping to open up a gap in the energy spectrum, and hence look for the existence of the topological phases. The semi-Dirac system with the Haldane term  has been discussed in literature \cite{mondal2021}. However to the best of our knowledge, the semi-Dirac system with the Haldane term and a real third neighbour hopping have never been  discussed and hence is new to the scientific community. An important dividend of such an exercise will be accessing regions in the phase diagram with large values of the Chern number, which also facilitates studying the topological phase transitions between phases with different Chern numbers.
	
	Motivated by the above scenario, here we discuss the topological properties of the semi-Dirac system in presence of a third neighbour (between different sublattices) hopping. We shall show that inclusion of the third neighbour hopping shifts the band minima from the  boundary towards the interior of the BZ and makes the system a Chern insulator, with Chern numbers $\pm2$ for certain values of the hopping amplitude. Addition of the Semenoff mass to the problem changes Chern number from $\pm2$ to $\mp1$. Consequently, we obtain the plateaus of the Hall conductivity that are quantized as $C e^2/h$, with $C$ being its Chern number and acquires values $\pm 1$ and $\pm 2$.
	
	
	The paper is organized as follows, in section \ref{sec:modelhamiltonian} we show the semi-Dirac Hamiltonian in presence of a Haldane term and a real third neighbour hopping on a honeycomb lattice. In section \ref{sec:phasediagram}, we investigate the topological properties by computing the Chern number for various values of the amplitude of the third neighbour hopping, and obtain the phase diagrams that demonstrate the existence (or absence) of  the non-trivial topological phases. In section \ref{sec:edgestates}, we study the structure of the edge modes in a nanoribbon for various relevant values of the second and the third neighbour hopping amplitudes. Hence, we compute the anomalous Hall conductivities  in section \ref{sec:hallcond} that exhibit plateaus quantized in units of $e^2/h$ and finally conclude with a brief summary of our results in section \ref{sec:conclusion}.
	
	\section{Model Hamiltonian}\label{sec:modelhamiltonian}
\begin{figure}[h]
		\centering
		\includegraphics[width=0.4\textwidth]{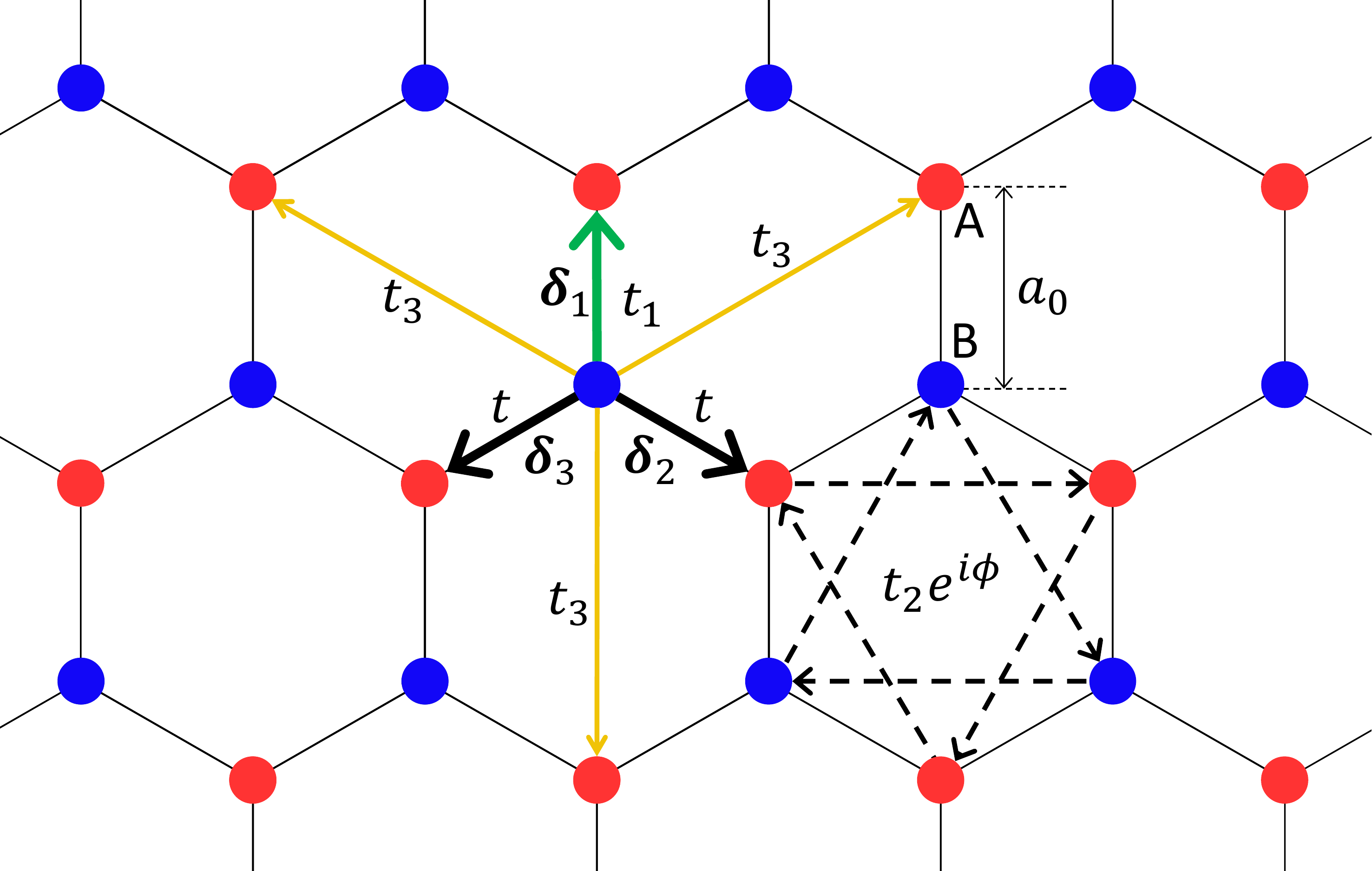}
		\caption{\raggedright A honeycomb lattice is shown where the red and the blue circles represent the sublattices A and B respectively. In the $\boldsymbol{\delta}_2$ and $\boldsymbol{\delta}_3$ directions the N2 hopping strengths are same ($t$), while in the $\boldsymbol{\delta}_1$ direction it is $t_1$. The N3 hopping  is shown by the yellow arrow.}\label{honeycomb}
	\end{figure}

	We consider a tight binding Hamiltonian on a honeycomb lattice with hopping between the various neighbours that can be written as, 
	
	\begin{align}\label{ham1}
		H =&- \sum_{\left<i,j\right>} t_{ij} c_i^{\dagger} c_j + t_2 \sum_{\left<\left<i, j\right>\right>} e^{i\phi_{ij}} c_i^{\dagger} c_j + \nonumber\\ 
		& + t_3\sum_{\left<\left<\left<i,j\right>\right>\right>}c_i^{\dagger} c_j + \sum_i \Delta_i c_i^\dagger c_i + {\rm {h.c.}}
	\end{align}
	The first term is the nearest neighbour (N1) hopping. The N1 hopping strengths along the $\boldsymbol{\delta}_2$ and the $\boldsymbol{\delta}_3$ directions are $t$, while in the third direction, that is along $\boldsymbol{\delta}_1$, the strength is $t_1$ as shown in Fig. \ref{honeycomb}. The N1 vectors are given by $\boldsymbol{\delta}_1 = a(0, 1)$, $\boldsymbol{\delta}_2= a(\sqrt{3}/2, -1/2)$ and $\boldsymbol{\delta}_3= a(-\sqrt{3}/2, -1/2)$. In the first term, $t_{ij} = t_1$ or $t$ when $j$ connects the neighbours $\boldsymbol{\delta}_{1}$ or $\boldsymbol{\delta}_{2,3}$ that belongs to the other sublattice respectively.  We have assumed two different values of  $t_1$, such as, $t_1 = t$ and $t_1 = 2t$. The value $t_1 = t$ represents the well known isotropic Dirac case, such as graphene, while $t_1=2t$ denotes the semi-Dirac case and is the focus for this work. The second term is the Haldane term comprising of a complex second neighbour (N2) hopping with an amplitude, $t_2$ and  a complex phase denoted by $\phi_{ij}$, where $\phi$ assumes positive (negative) values if the electron hops in the clockwise (anti-clockwise) direction. The third term represents the third neighbour (N3) hopping between different sublattices and the fourth term represents the onsite energy (Semenoff mass), that assumes values $+\Delta$ and $-\Delta$ for sublattices A and B respectively. Performing a Fourier transform of  Eq. \ref{ham1}, the Hamiltonian in the momentum space can be written as,
	
	\begin{align}\label{ham_kspace}
		H(\mathbf{k})&= h_x(\mathbf{k}) \sigma_x + h_y(\mathbf{k}) \sigma_y + h_z(\mathbf{k}) \sigma_z + h_0(\mathbf{k}) I \nonumber\\
		& = \mathbf{h(\mathbf{k})}\cdot\boldsymbol{\sigma} + h_0(\mathbf{k})I
	\end{align}
	where,
	\begin{align}
		h_x(\mathbf{k}) =& \left\{ t_1 \cos k_y + 2t\cos \frac{k_y}{2} \cos\frac{\sqrt{3}k_x}{2} \right\} + \nonumber\\
		&t_3\left\{ \cos 2k_y + 2\cos k_y \cos \sqrt{3}k_x \right\},
	\end{align}
	\begin{align}
		h_y(\mathbf{k}) =& \left\{ -t_1 \sin k_y + 2t\sin \frac{k_y}{2} \cos\frac{\sqrt{3}k_x}{2} \right\} + \nonumber\\
		&t_3\left\{ \sin 2k_y - 2\sin k_y \cos \sqrt{3}k_x \right\},
	\end{align}
	\begin{align}
		h_z(\mathbf{k})= M-2t_2\sin \phi\left\{2\sin\frac{\sqrt{3}k_x}{2} \cos\frac{3k_y}{2} - \sin\sqrt{3}k_x \right\}
	\end{align}
and, 
\begin{align}
	h_0(\mathbf{k}) = 2t_2\cos \phi\left\{2\cos\frac{\sqrt{3}k_x}{2} \cos\frac{3k_y}{2} + \cos\sqrt{3}k_x \right\}
\end{align}
where $\sigma_i$ ($i\in x,y,z$) denote the 2$\times$2 spin-1/2 Pauli matrices which  represent the sublattice degrees of freedom, and $I$ is the 2$\times$2 identity matrix. The energy dispersion can be obtained as, 
\begin{equation}
E(\mathbf{k}) = h_0(\mathbf{k}) + \sqrt{h_x(\mathbf{k})^2 + h_y(\mathbf{k})^2 + h_z(\mathbf{k})^2},	
\end{equation}
where the $\pm$ signs refer to the upper (conduction) band and the lower (valence) band respectively. In the absence of $t_2$ and $t_3$, the band dispersion is linear along one direction and quadratic along its perpendicular direction \cite{sinha2020} about the band touching $\mathbf{M}$ point in the BZ. We refer to this as the zero mode in our subsequent discussion.

\begin{figure}[h]
		\centering
		\begin{subfigure}[b]{0.23\textwidth}
			\includegraphics[width=\textwidth]{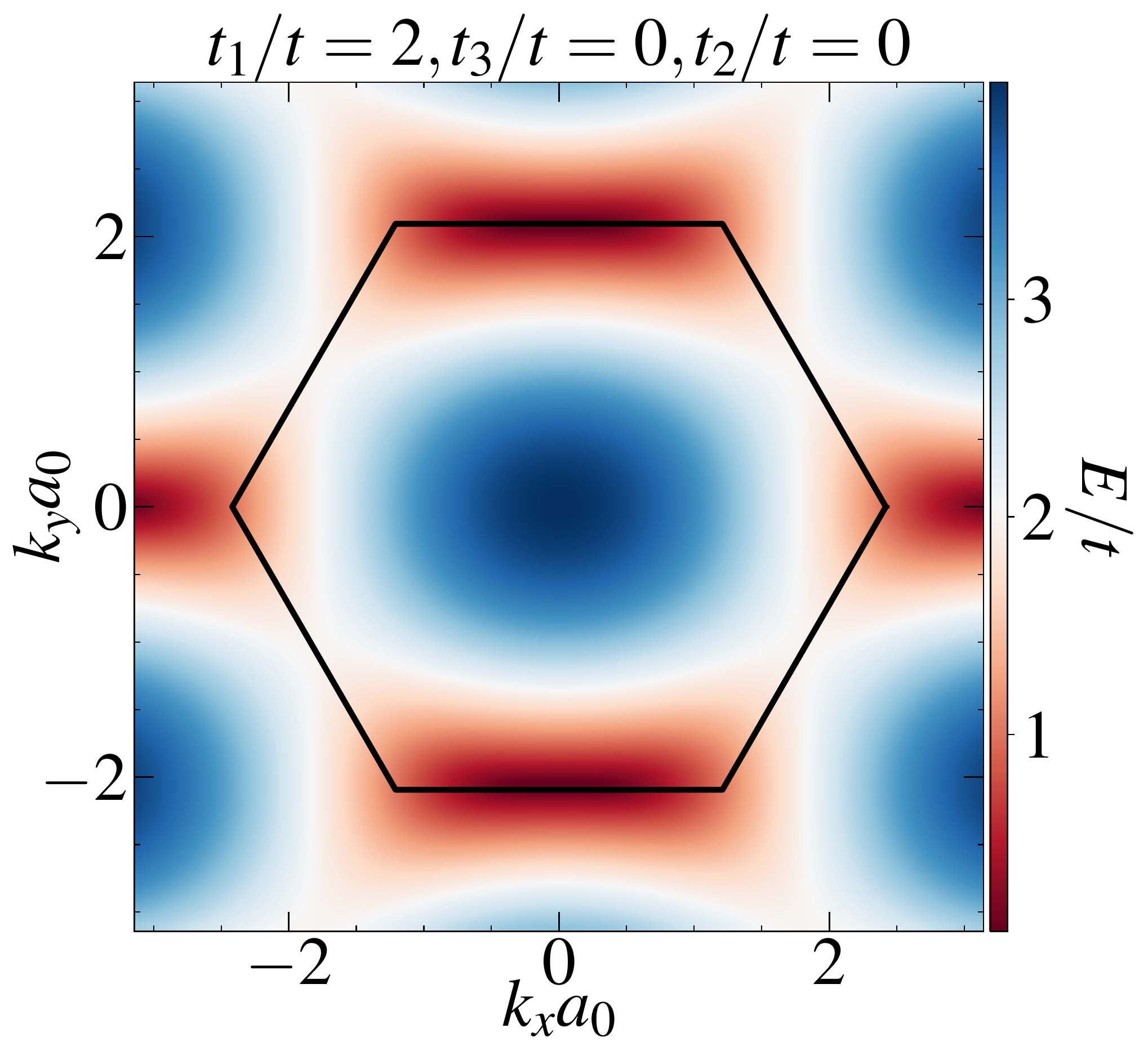}
			\subcaption{}\label{fig:bt_t1_2_t2_0_t3_0_m_0}
		\end{subfigure}
		\begin{subfigure}[b]{0.23\textwidth}
			\includegraphics[width=\textwidth]{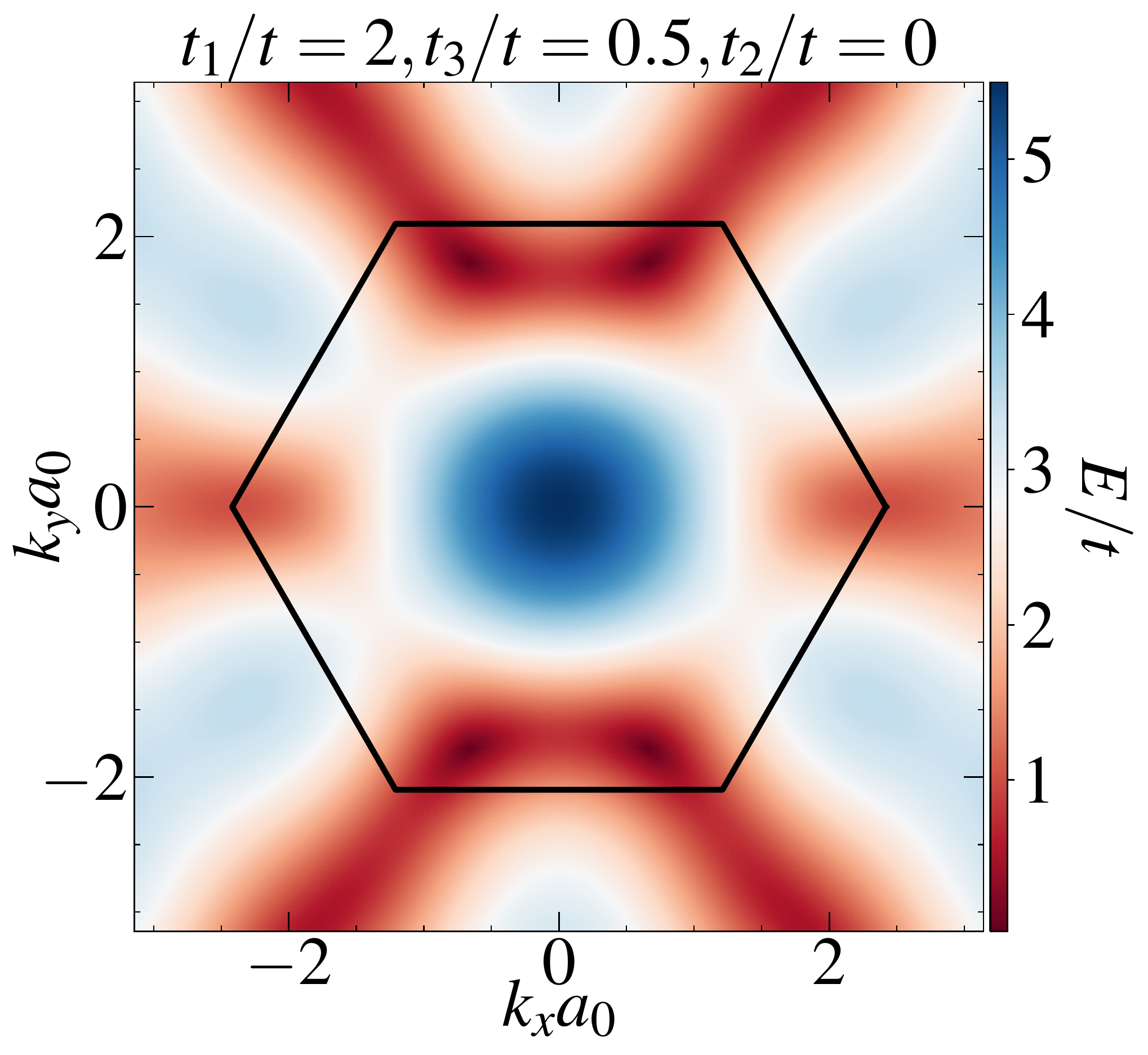}
			\subcaption{}\label{fig:bt_t1_2_t2_0_t3_0.5_m_0}
		\end{subfigure}
		\begin{subfigure}[b]{0.23\textwidth}
			\includegraphics[width=\textwidth]{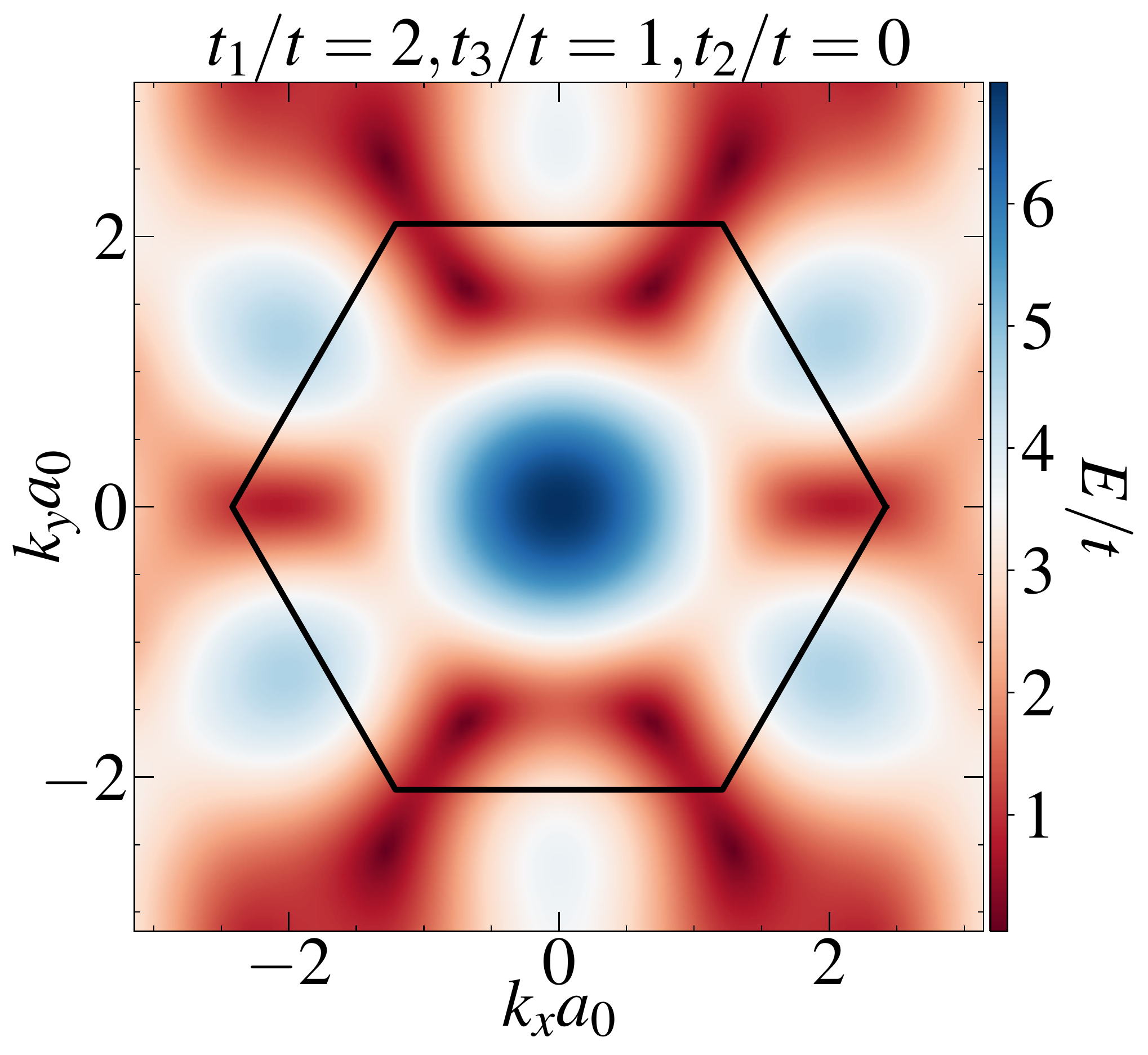}
			\subcaption{}\label{fig:bt_t1_2_t2_0_t3_1_m_0}
		\end{subfigure}
		\begin{subfigure}[b]{0.235\textwidth}
			\includegraphics[width=\textwidth]{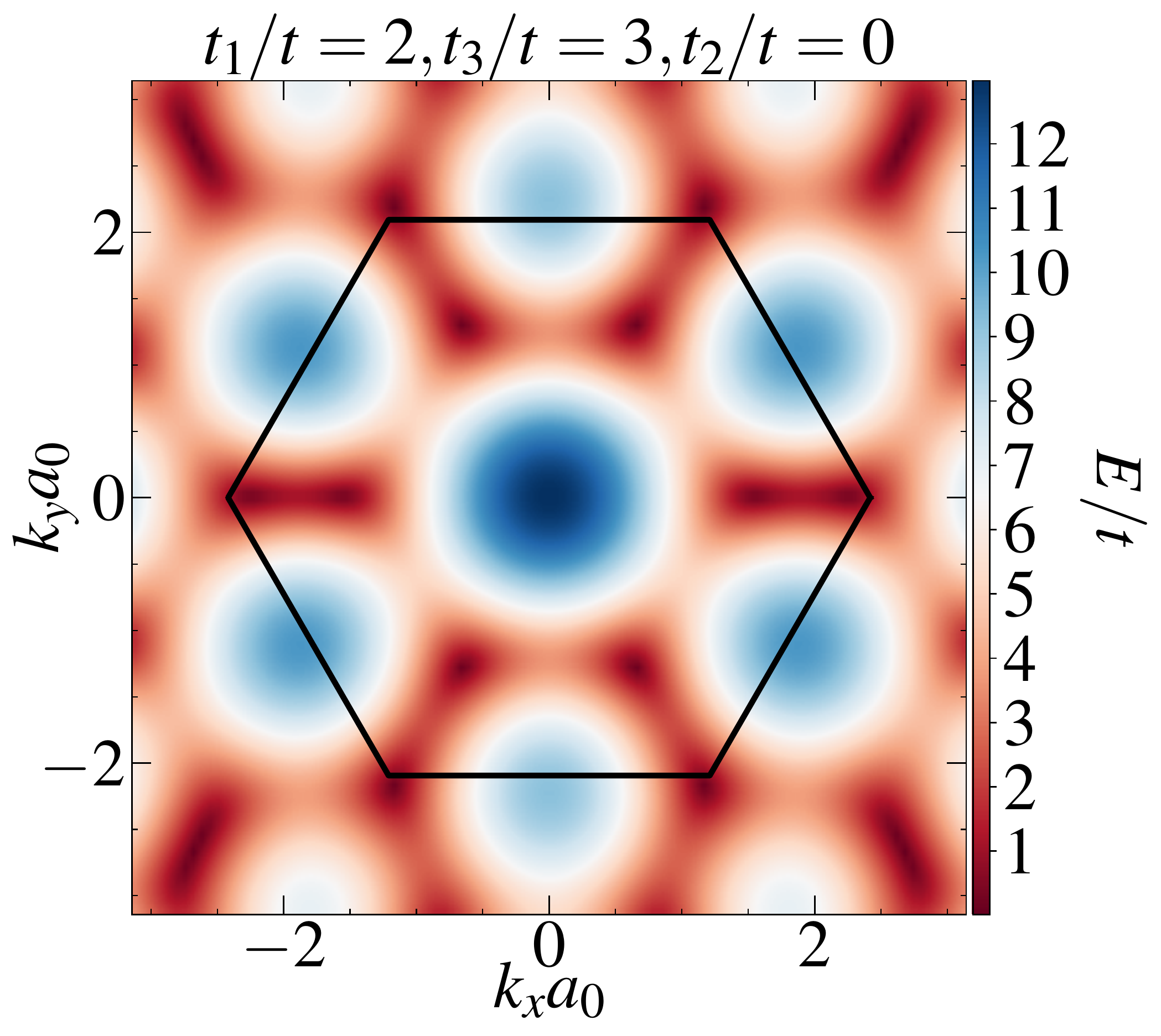}
			\subcaption{}\label{fig:bt_t1_2_t2_0_t3_3_m_0}
		\end{subfigure}
		\begin{subfigure}[b]{0.23\textwidth}
			\includegraphics[width=\textwidth]{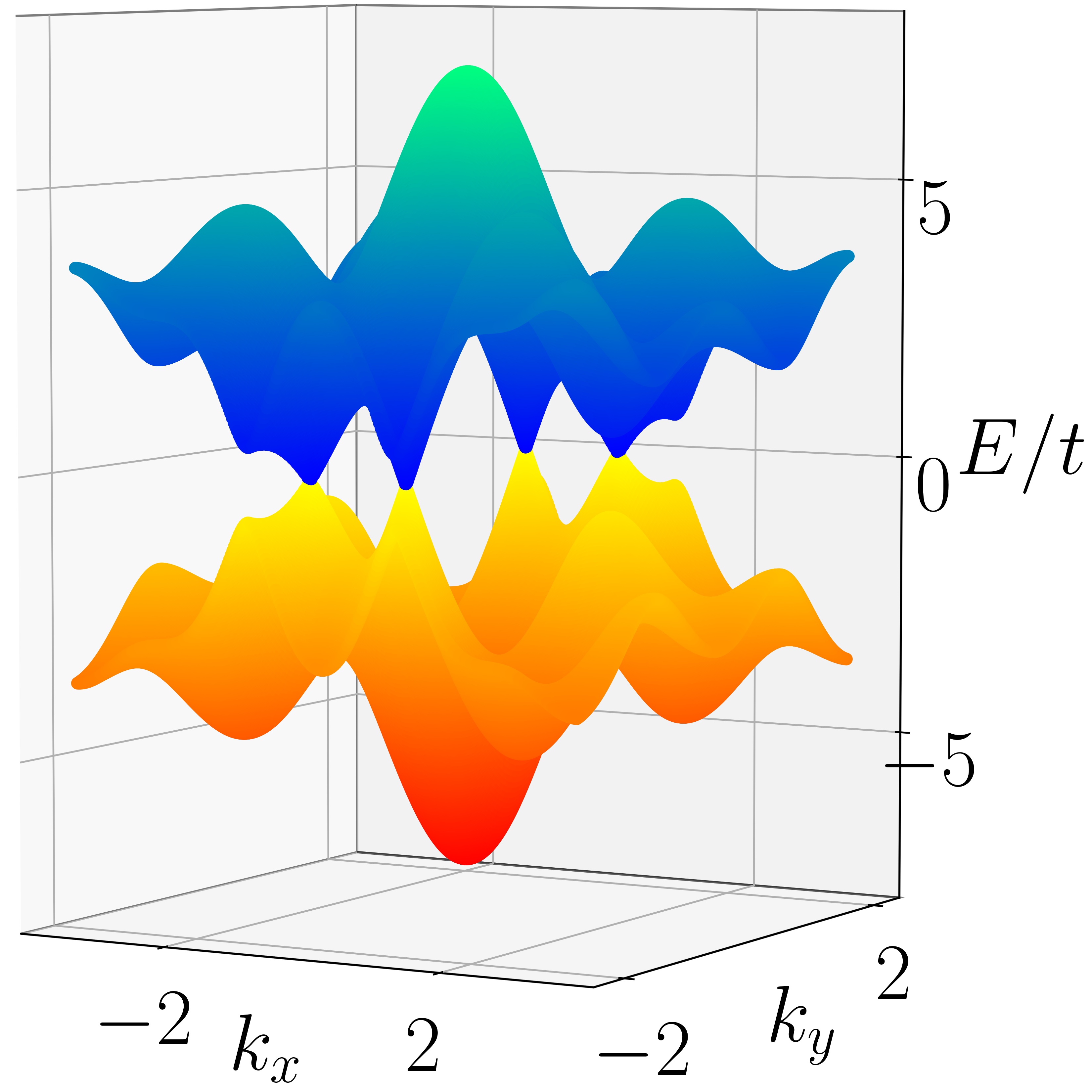}
			\subcaption{}\label{fig:band1}
		\end{subfigure}
		\begin{subfigure}[b]{0.23\textwidth}
			\includegraphics[width=\textwidth]{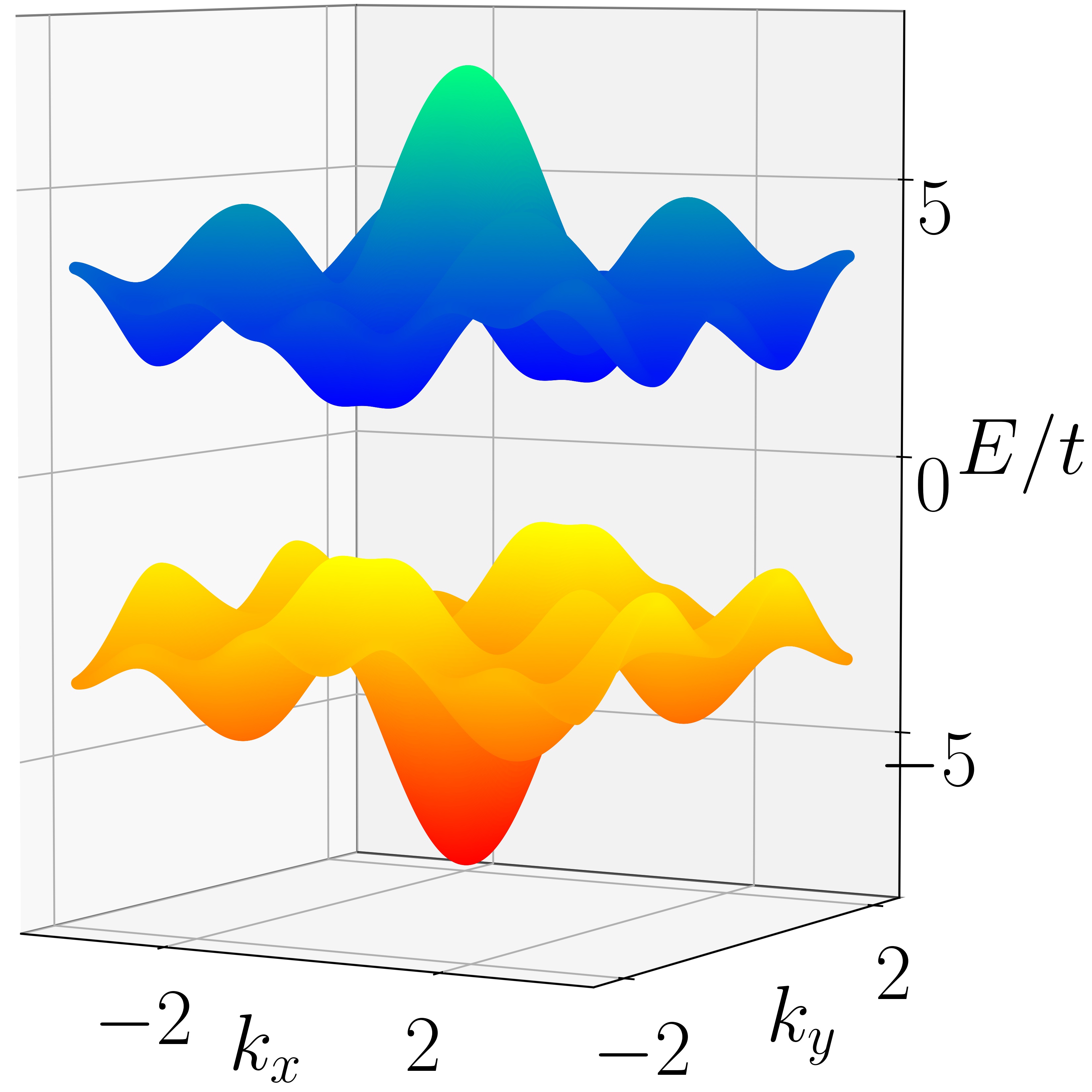}
			\subcaption{}\label{fig:band2}
		\end{subfigure}
		\caption{\raggedright The top view of the bandstructure for the semi-Dirac ($t_1 = 2t$) system is depicted for (a) $t_3/t = 0$, (b) $t_3/t = 0.5$, (c) $t_3/t = 1$ and (d) $t_3/t = 3$. The hexagons in each figure represent the first Brillouin zone. In the calculations, we have fixed $t_2 = 0$, $\Delta = 0$. In (e) and (f) a three dimensional depiction of the bandstructure for the semi-Dirac system are shown for $t_2 = 0$ and $t_2 = 0.5t$ respectively, where we have used $\Delta = 0$ and $\phi = \pi/2$.}
		\label{fig:band}
	\end{figure}

Now, if we add  a small N3 hopping, namely, $t_3$, then the zero modes shift from the $\mathbf{M}$ point towards the interior of the BZ as shown in Figs. \ref{fig:bt_t1_2_t2_0_t3_0.5_m_0}-\ref{fig:bt_t1_2_t2_0_t3_3_m_0}. There are four zero modes inside the first BZ for a non-zero value of $t_3$. Let us call these points where the zero modes occur as $\boldsymbol{\Lambda}$ points. For example, one of the zero modes for a particular value of $t_3$, namely, $t_3 = t$ occurs approximately at a particular $\boldsymbol{\Lambda}$ point, namely, $\boldsymbol{\Lambda}_1$ = $\left(\frac{2.0023\pi}{9a_0}, \frac{1.5487\pi}{3a_0}\right)$, while the same for a different value of $t_3$, namely $t_3 = 3t$ approximately occurs at another, $\boldsymbol{\Lambda}_2 = \left( \frac{1.0805\pi}{3\sqrt{3}a_0}, \frac{1.2403\pi}{3a_0} \right)$. For other values of $t_3$, namely, say $t_3>3t$, the zero modes remain fixed at the same locations as that for $t_3 = 3t$. 

Now, if we turn on the N2 hopping, $t_2$ then the spectral gaps open up at these $\Lambda$ points in the BZ where the zero modes occur, and hence the system behaves as an insulator. However, in the absence of $t_3$ (with $t_2$ being non-zero), there is no gap at the $\mathbf{M}$ point, and the dispersion is anisotropic linear (linear along both the directions, but with different velocities along the $x$- and  the $y$-directions) about the $\mathbf{M}$ point, which makes the system a semi-metal as discussed in Ref. \cite{mondal2021}. In Figs. \ref{fig:band1} and \ref{fig:band2}, we have shown the bandstructures for the semi-Dirac system in the absence, and in the presence of $t_2$ respectively for non-zero values of $t_3$. In our calculations, we have fixed the values of the Haldane flux, $\phi$, N1 hopping, $t_1$ and the Semenoff mass, $\Delta$ to be $\pi/2$, $2t$ and zero respectively. The corresponding bandstructures for the Dirac system have been discussed in Ref. \cite{sticlet2013,bena2011}, and we skip them here to make our discussion concise.

It may be noted that $t_3$ is indeed a parameter and the values used may not have experimental relevance. The reason being that the value of the real second neighbour hopping is of the order of 0.1eV \cite{castroneto}, which would mean that $t_3$ is even smaller. However, the phase diagrams presented in section \ref{sec:phasediagram} demands the value of $t_3$ to be of the order of $t$ or even larger in order to access topological phases with different Chern numbers.

	\section{The Phase diagram}\label{sec:phasediagram}
	In this section we obtain the phase diagram by numerically calculating the Chern number of the system. Since in this model, the complex N2 hopping term breaks the TRS, non-zero values and hence non-trivial phases with finite Chern number are expected. The inversion symmetry breaking onsite energies, $\pm\Delta$ on different sublattices open or close  energy gaps in the energy spectrum at the $\Lambda$ points. We compute the Chern number via \cite{thouless,avron1988},
	
	\begin{eqnarray}\label{eq:chern_number}
		C & = & \frac{1}{2\pi}\int\int_{\mathrm{BZ}}\Omega(k_x, k_y)\mathrm{d}k_x \mathrm{d}k_y,
	\end{eqnarray}
	where $\Omega(k)$ denotes the $z$-component of the Berry curvature which is given by,
	
	\begin{eqnarray}\label{eq:berry_curv}
		\Omega(k_x, k_y) = \frac{\mathbf{h}}{2|\mathbf{h}|^3}\cdot\left( \frac{\partial \mathbf{h}}{\partial k_x} \times \frac{\partial \mathbf{h}}{\partial k_y} \right)
	\end{eqnarray}
	
	\begin{figure}[h]
		\centering
	\begin{subfigure}[b]{0.23\textwidth}
		\includegraphics[width=\textwidth]{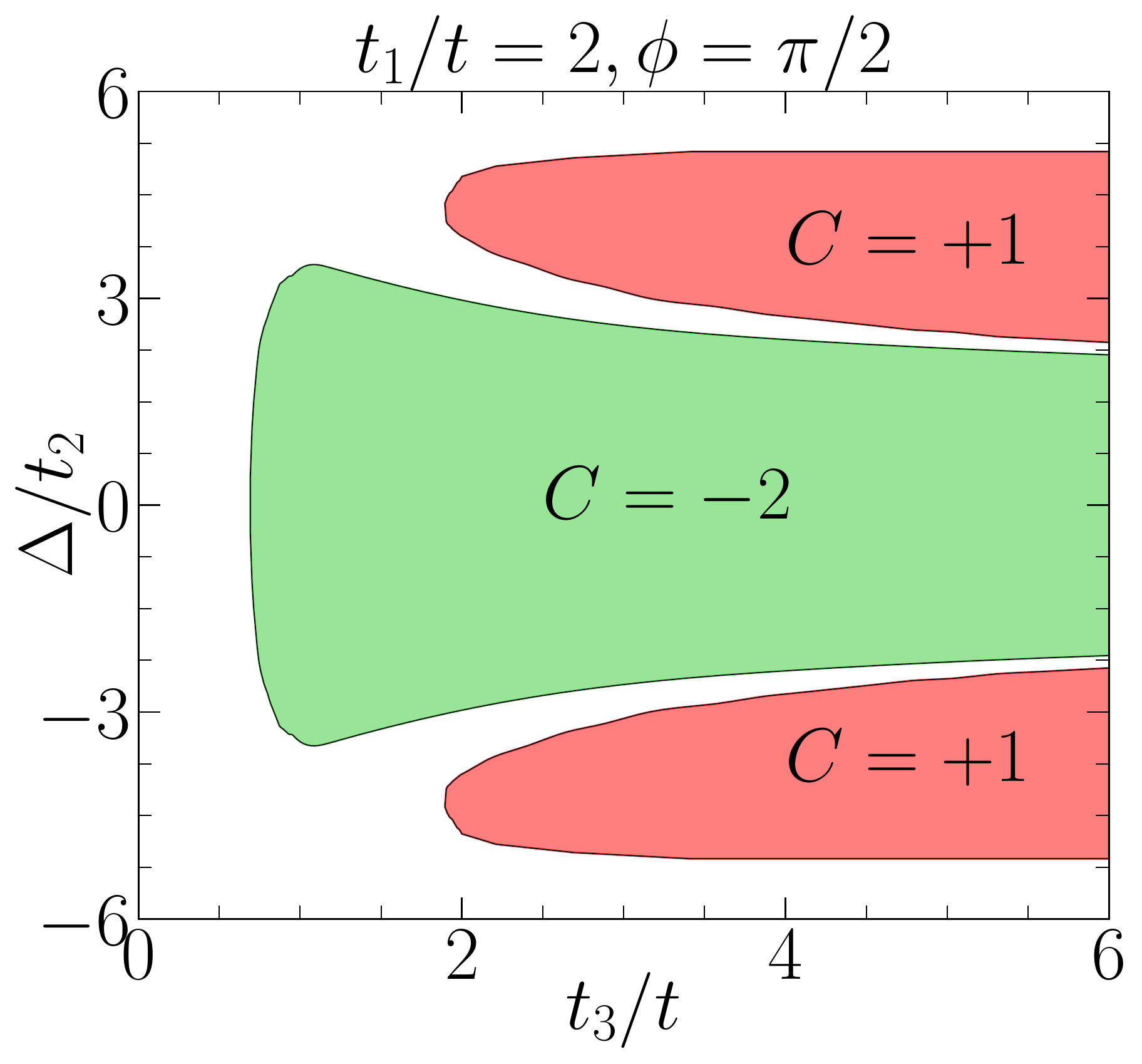}
		\subcaption{}\label{fig:pd_m_t3_t1_2}
	\end{subfigure}
	\begin{subfigure}[b]{0.23\textwidth}
		\includegraphics[width=\textwidth]{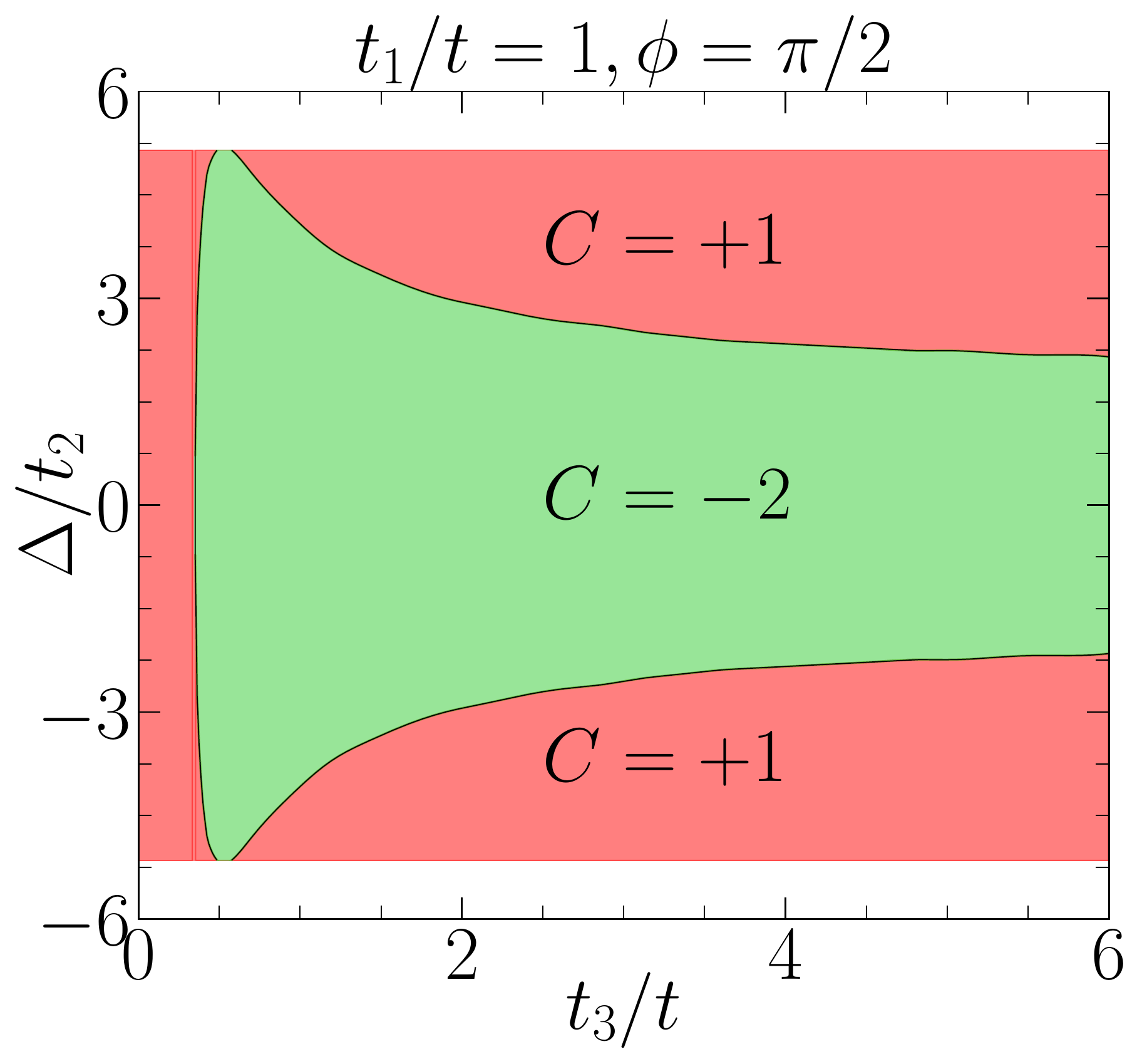}
		\subcaption{}\label{fig:pd_m_t3_t1_1}
	\end{subfigure}
	\begin{subfigure}[b]{0.23\textwidth}
		\includegraphics[width=\textwidth]{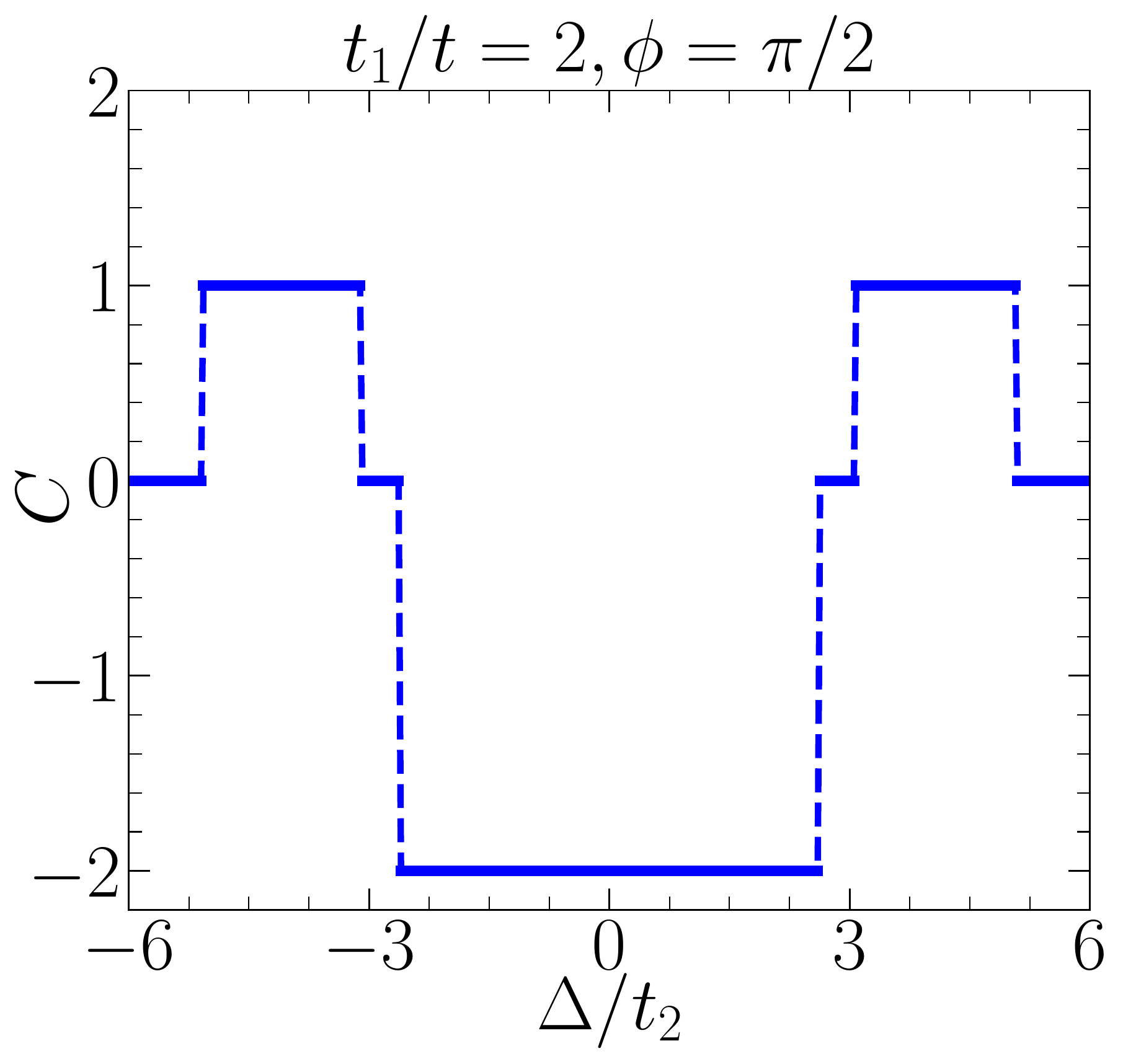}
		\subcaption{}\label{fig:C_vs_m_t1_2}
	\end{subfigure}
	\begin{subfigure}[b]{0.23\textwidth}
		\includegraphics[width=\textwidth]{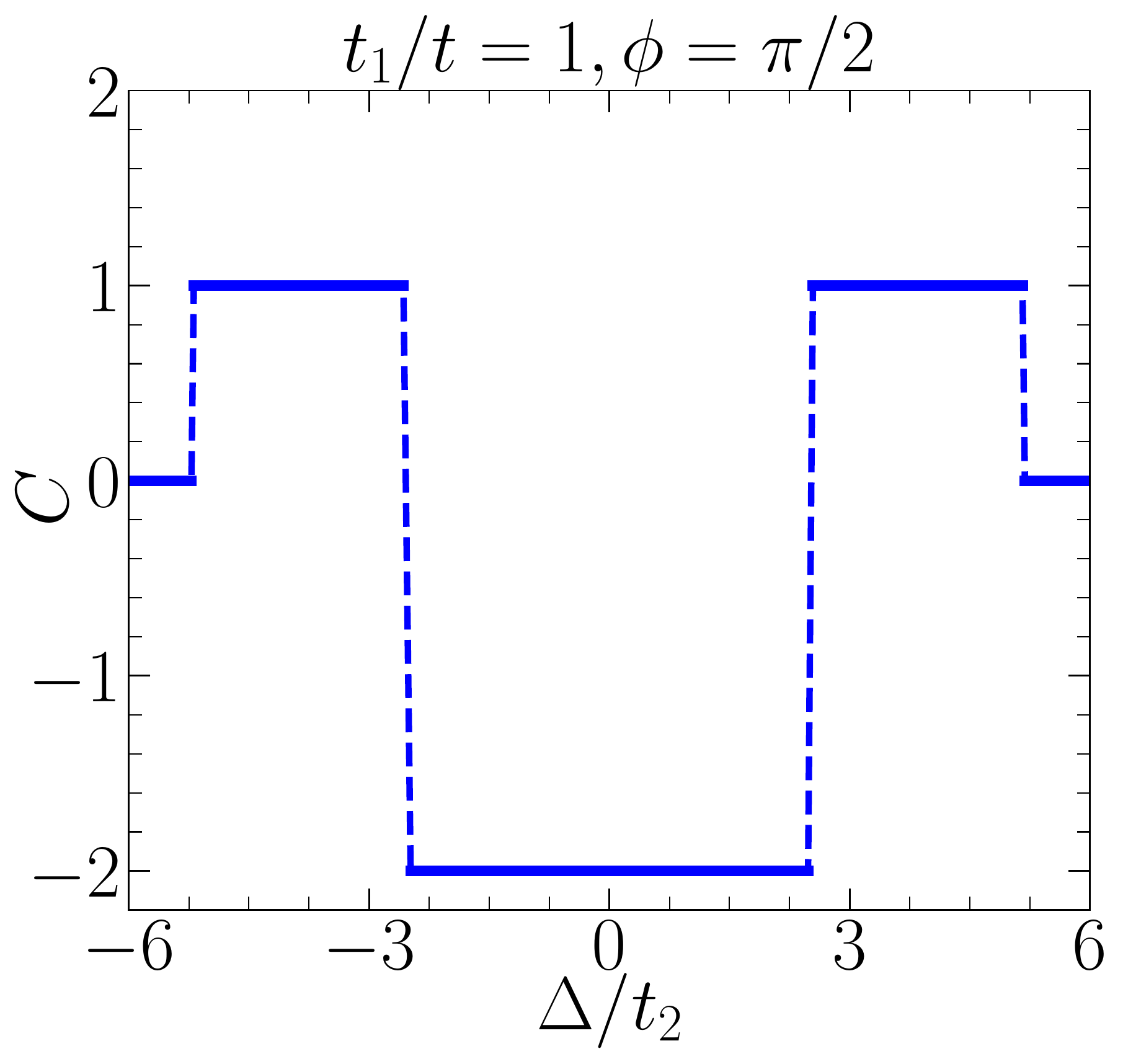}
		\subcaption{}\label{fig:C_vs_m_t1_1}
	\end{subfigure}
		\caption{\raggedright The Chern number of the lower band is depicted as a function of $\Delta$ and $t_3$ for (a) $t_1/t = 2$ and (b) $t_1/t = 2$. The coloured regions signify the Chern insulating regions with non-zero Chern numbers ($C = +1$ for the red region and $C = -2$ for the green one), while the white region denotes the trivial insulating phase with $C=0$. We have shown the variation of $C$ as a function of $\Delta$ for a particular value of $t_3$, say $t_3=3t$, in (c) and (d) for $t_1/t=2$ and $t_1/t = 1$ respectively. In this calculation, the Haldane flux $\phi$ is kept fixed at $\pi/2$. The topological phase transitions are implied via $C$ discontinuously changing values between $1\rightarrow 0 \rightarrow-2$.}
		\label{fig:pd_m_t3}
	\end{figure}
	\begin{figure}[h]
			\begin{subfigure}[b]{0.32\textwidth}
				\includegraphics[width=\textwidth]{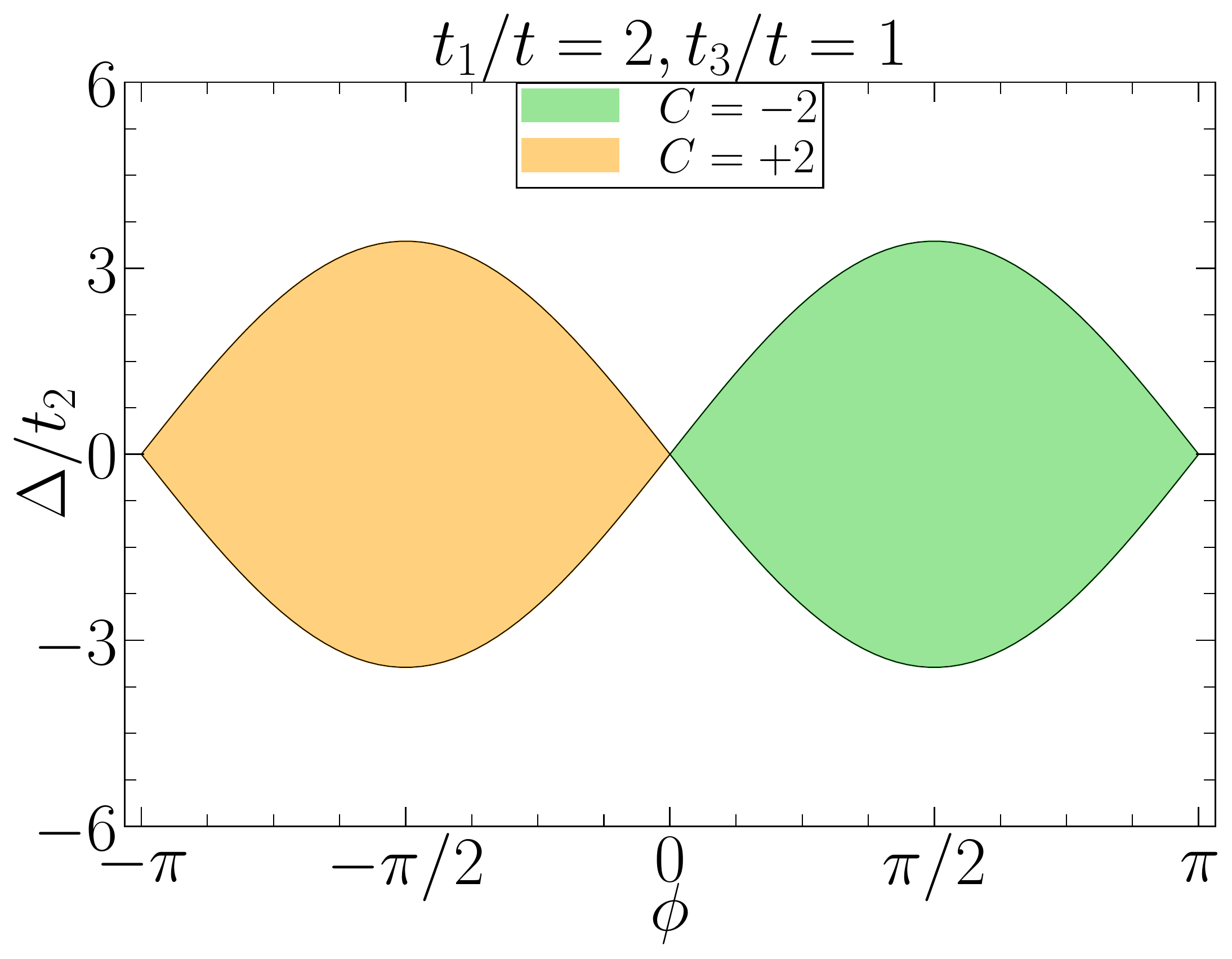}
				\subcaption{}\label{fig:pd_m_phi_t1_2_t3_1}
			\end{subfigure}
			\begin{subfigure}[b]{0.32\textwidth}
				\includegraphics[width=\textwidth]{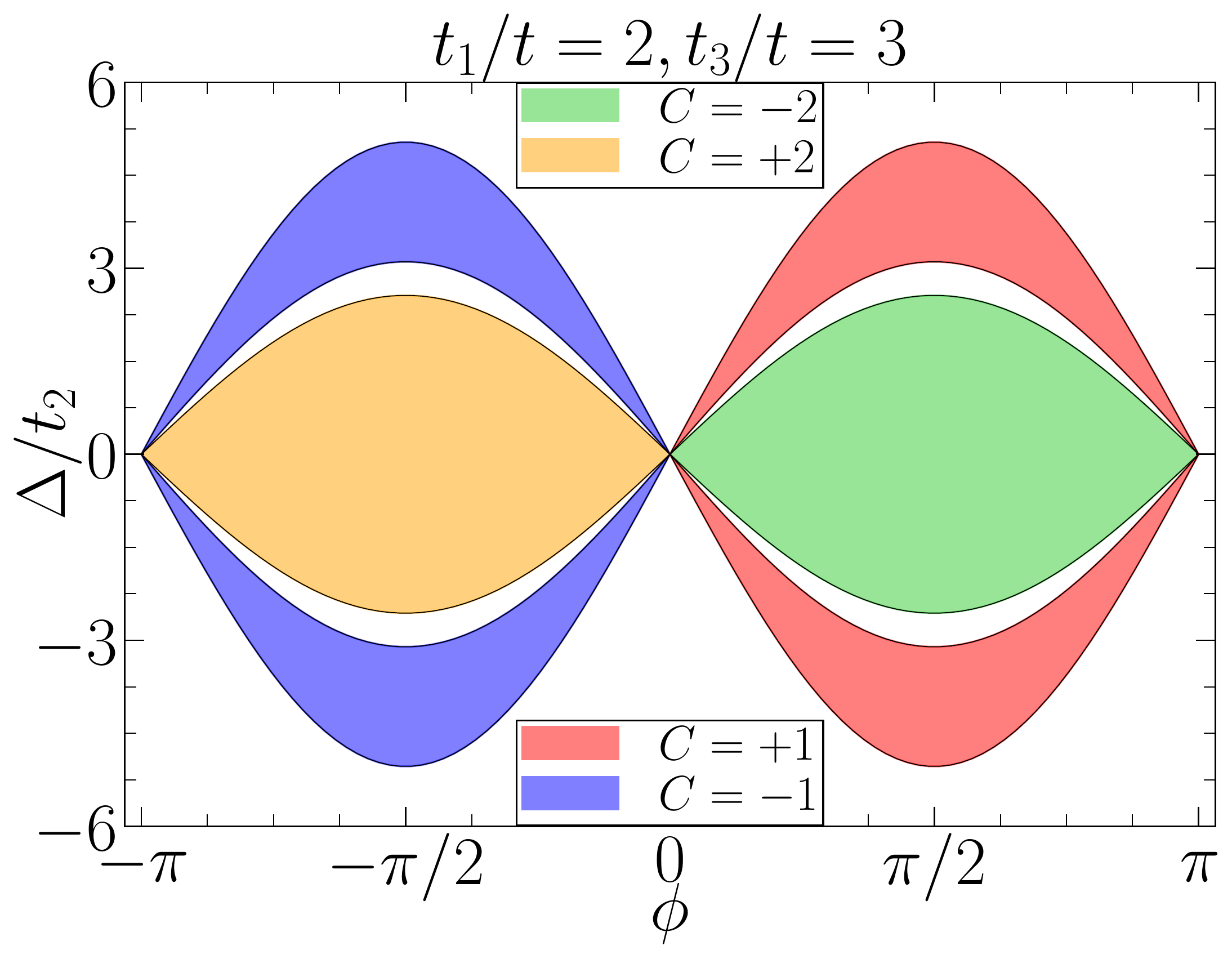}
				\subcaption{}\label{fig:pd_m_phi_t1_2_t3_3}
			\end{subfigure}
			\begin{subfigure}[b]{0.32\textwidth}
				\includegraphics[width=\textwidth]{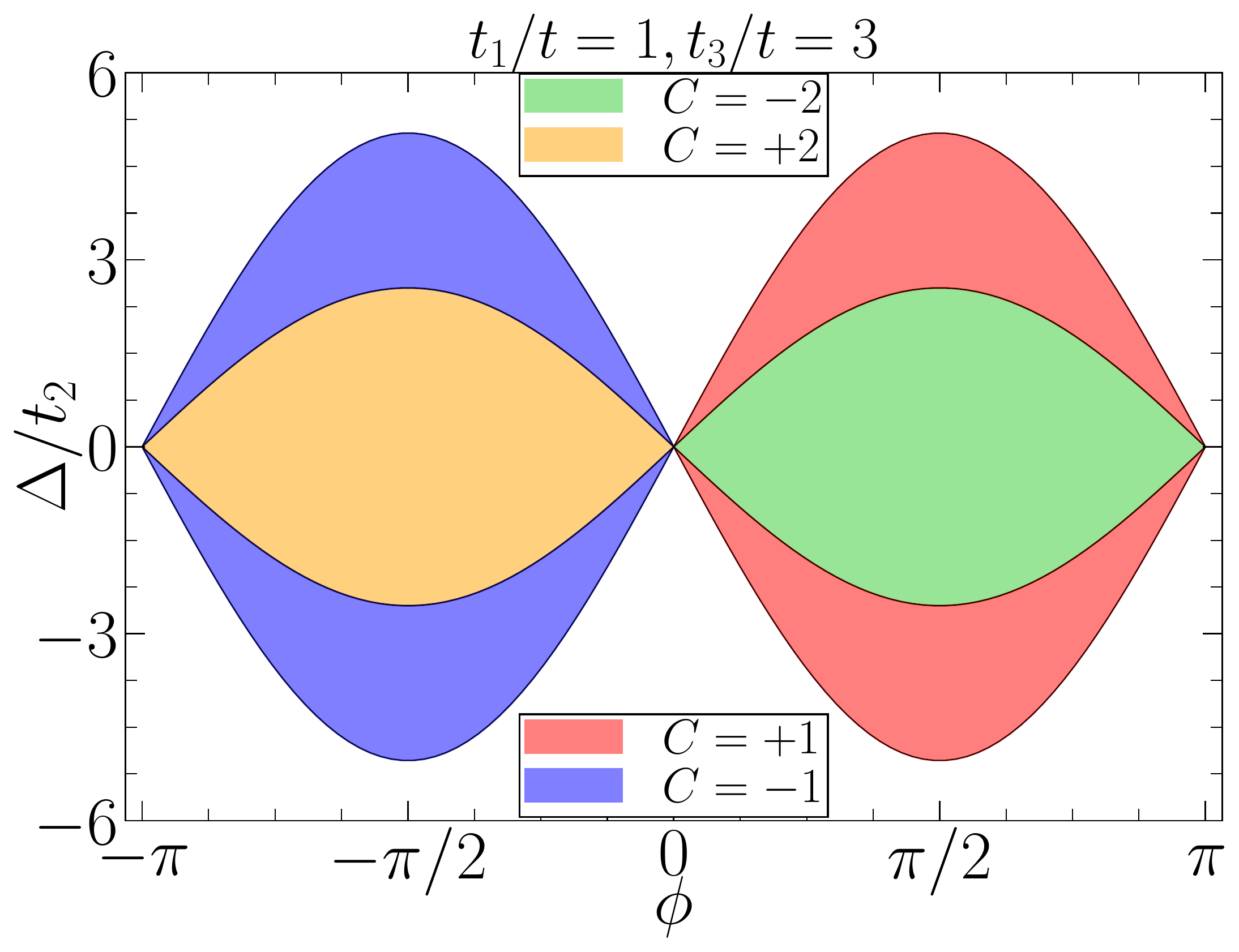}
				\subcaption{}\label{fig:pd_m_phi_t1_1_t3_3}
			\end{subfigure}
			\caption{\raggedright The Chern numbers of the lower band is shown as a function of $\Delta$ and $\phi$ for (a) $t_1/t = 2$ and $t_3/t = 1$, (b) $t_1/t = 2$ and $t_3/t = 3$ and (c) $t_1/t = t$ and $t_3/t = 3$. In each figure, the green and the orange regions denote the Chern insulating phase with the Chern numbers -2 and +2 respectively, while the red and the blue regions imply the Chern numbers +1 and -1 respectively. Further, the white region denotes trivial topological regime with zero Chern number.}
			\label{fig:pd_m_phi}

	\end{figure}
 	\noindent where $\mathbf{h(\mathbf{k})}$ is defined in Eq. \ref{ham1}. It is to be noted that in the absence of an N3 hopping, the Chern number is always zero for any arbitrary value of $\Delta$ and $\phi$ for the semi-Dirac case ($t_1 =2t$), even though the time reversal symmetry remains broken. However in presence of the non-zero N3 hopping, we may obtain non-zero values for the Chern number.
	In Fig. \ref{fig:pd_m_t3_t1_2} we have depicted the Chern number corresponding to the lower band as a function of $\Delta$ and $t_3$ for Haldane flux, $\phi=\pi/2$. As can be seen from Fig. \ref{fig:pd_m_t3_t1_2}, there are two regions denoted by the red and the green colours. The region in red indicates the value of the Chern number, $C = 1$, while the green region indicates $C = -2$. In addition, there is also a finite region denoted by the white colour, which corresponds to a trivial region with $C=0$. In absence of $t_3$ or at small values of $t_3$, namely $t_3<0.68t$, for all values of $\Delta$, the trivial region prevails.
	We observe the topological phase with the Chern number $C=-2$ beyond a certain value of the N3 hopping, $t_3$, namely $t_3\gtrsim0.68t$, for a zero Semenoff mass ($\Delta = 0$). If we increase the value of $\Delta$, then we observe the $C= -2$ phase for a range of values of $t_3$, such as, $0.68t\lesssim t_3 \lesssim 1.9t$. However, for $t_3 \gtrsim 1.9t$, there are two topological phases with Chern numbers $C = 1$ and $C=-2$, which depend on the value of $\Delta$. For example, for $t_3 = 3t$, there is phase transition occurring from a $C=0$ to a $C=1$ phase at $\Delta \simeq-5.04 t_2$. $C$ again drops to zero at $\Delta\simeq -3.11t_2$. Beyond $\Delta\simeq-2.56t_2$ the Chern number becomes $-2$ and stays at $-2$ until $\Delta \simeq 2.56t_2$, when the Chern number vanishes again. The Chern number becomes 1 at $\Delta\simeq3.11t_2$, and finally vanishes again for $\Delta\gtrsim 5.04 t_2$. Thus there are a series of phase transitions occurring at $t_3 = 3t$. There is always a trivial region (with $C = 0$) in between the two Chern insulating regions having two different Chern numbers (the white region between the red and the green regions). Further, as one increases the value of $t_3$, the vanishing Chern numbers are obtained for lesser values of $\Delta$. As a result, the width of the Chern insulating region with $C=-2$ (the green region) shrinks with the increase of $t_3$, or equivalently, we can say, the width of the $C=1$ region increases with increase in $t_3$. The trivial region (shown in white in Fig. \ref{fig:pd_m_t3_t1_2}) gets narrower as one increases the value of $t_3$. 
		
	This phenomenon is somewhat different in the Dirac case (see Fig. \ref{fig:pd_m_t3_t1_1}), where we can see a non-zero Chern number (namely, $C = 1$) even in absence of the N3 hopping, that is, the Haldane model. The phase persists for very small values of $t_3$. However in the  presence of N3 hopping, we obtain a phase with Chern number $C=-2$ or $C = 1$ depending on the value of $\Delta$. Further, unlike the semi-Dirac case, there is no trivial regime in between the two different Chern insulating regimes, that is, the red and the green regions. If we fix the value of $t_3$, say, $t_3 = 3t$, and calculate the Chern numbers for increasing values of $\Delta$, then we observe the Chern number to jump from $C=-2$ to $C=1$ at $\Delta \simeq 2.55t_2$. Finally, the Chern number drops to zero from a value $C=1$ at $\Delta = 3\sqrt{3}t_2$. The values of $\Delta$, at which the Chern number changes from a value $C=-2$ to $C=1$, depend on the value of the N3 hopping $t_3$ (see the shoulder like region in Fig. \ref{fig:pd_m_t3_t1_1}). However, the values of $\Delta$ at which the Chern number vanishes from a value $C=1$ does not depend upon $t_3$. It should be noted that the calculations are done for a Haldane flux, $\phi=\pi/2$. If we change $\phi$ to $-\pi/2$ then the phase diagram will remain identical, except that the Chern numbers will undergo a sign change.
	
	In Figs. \ref{fig:C_vs_m_t1_2} and \ref{fig:C_vs_m_t1_1} we have shown the variation of Chern numbers as a function of $\Delta$ for the semi-Dirac ($t_1 = 2t$) and the Dirac systems ($t_1=t$) respectively for a particular value of $t_3$, say, $t_3 = 3t$. As can be seen for the semi-Dirac case (see fig. \ref{fig:C_vs_m_t1_2}), there are phase transitions occurring from $C=0$ to $C = 1$ and then again to $C= 0$ as one increases $\Delta$. With further increase of $\Delta$, $C$ drops to $-2$.  To quote some numerical values, the plateau at $C = -2$ exists for a range of $\Delta$, that is, $-2.56t_2\lesssim\Delta \lesssim2.56t_2$. With  further increase in the value of $\Delta$, $C$ drops to zero and then again rises to 1 and finally vanishes. The plateaus at $C = 1$ persist for some values of $\Delta$, such that, $-5.04t_2 \lesssim \Delta \lesssim -3.11 t_2$ and $3.11t_2 \lesssim \Delta \lesssim 5.04 t_2$. A similar phase transitions are observed for the Dirac case (see Fig. \ref{fig:C_vs_m_t1_1}), except that there is direct phase transitions from $C = 1$ to $C = -2$ or vice versa. The $C = -2$ plateau occurs for $-2.55t_2 < \Delta < 2.55 t_2$, while the plateaus at $C = 1$ occur for $-3\sqrt{3}t_2 \leq \Delta \lesssim -2.55 t_2$ and $2.55t_2 \lesssim \Delta \leq 3\sqrt{3}t_2$.

	Fig. \ref{fig:pd_m_phi} shows the phase diagram in the $\Delta$-$\phi$ plane corresponding to the lower band for both the semi-Dirac and the Dirac cases. For both of them, the values of the Chern number depend on the value of N3 hopping amplitude (see Fig. \ref{fig:pd_m_t3}). In Fig. \ref{fig:pd_m_phi_t1_2_t3_1}, we have shown the phase diagram for the semi-Dirac case for $t_3 = t$. As can be seen, there are two Chern insulating regions with Chern numbers $C = -2$ (green region) and $C = +2$ (yellow region). The phase diagram is similar to that of the Haldane model, except that the values for the Chern number are different in this case. Further, the widths of the Chern insulating lobes are smaller than those in the Haldane model. Now, if we increase the value of $t_3$ (say, $t_3 = 3t$), we shall see additional Chern insulating regions emerge, with the Chern numbers given by $C = +1$ (red region) and $C = -1$ (blue region) as depicted in Fig. \ref{fig:pd_m_phi_t1_2_t3_3}. There exists a trivial insulating phase with $C = 0$ in between the two Chern insulating regions, that is, between the green and the red regions, or between the yellow and the blue regions. These types of phase diagrams are in complete contrast with the Dirac case, where in the latter, the trivial insulating phase is absent as shown in Fig. \ref{fig:pd_m_phi_t1_1_t3_3}. The width of the Chern insulating region with $C = -2$ is greater in the Dirac case compared to that of the semi-Dirac case (see Fig. \ref{fig:pd_m_phi_t1_1_t3_3}), as is evident from the $\Delta$-$t_3$ phase diagram (Fig. \ref{fig:pd_m_t3_t1_1}). For $t_3$ to be vanishingly small, we note that the phase diagram becomes similar to that of the Haldane model.
	
	The phase diagrams presented in Figs. \ref{fig:pd_m_t3} and \ref{fig:pd_m_phi} aid us in identifying specific values of $t_3$ and $\Delta$ to explore he nature of the topological phases. We achieve that via the numerical computation of the edge states and the anomalous Hall conductivity as discussed below. These quantities are investigated for $t_3 = 0.5t$, $t$ and $3t$, where we have considered $\Delta=0$ corresponding to $t_3 = 0.5t$ and $t$, while the $t_3=3t$ case has been studied for $\Delta = 0$ and $\Delta = 4t_2$, which, correspond to $C = -2$ and $C = 1$ respectively.
	
	\begin{figure}
		\includegraphics[width=\linewidth]{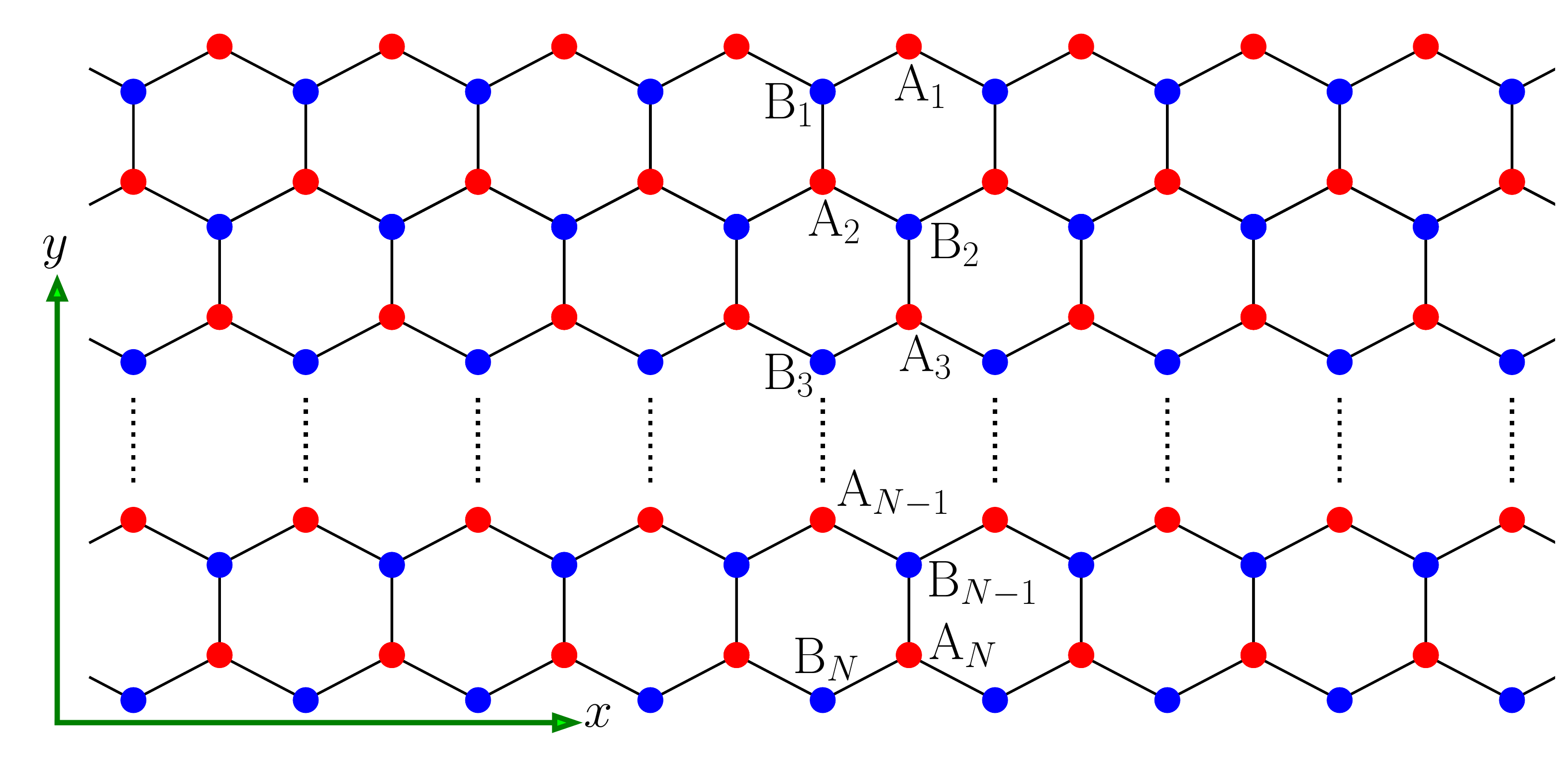}
		\caption{\raggedright A schematic diagram of the ribbon is shown. The edge current flows along the zigzag edges ($x$ axis) of the ribbon.}
		\label{fig:edge_ribbon}
	\end{figure}
	
	\begin{figure}[h]
		\captionsetup[subfigure]{labelformat=nocaption}
		\centering
		\includegraphics[width=0.85\linewidth]{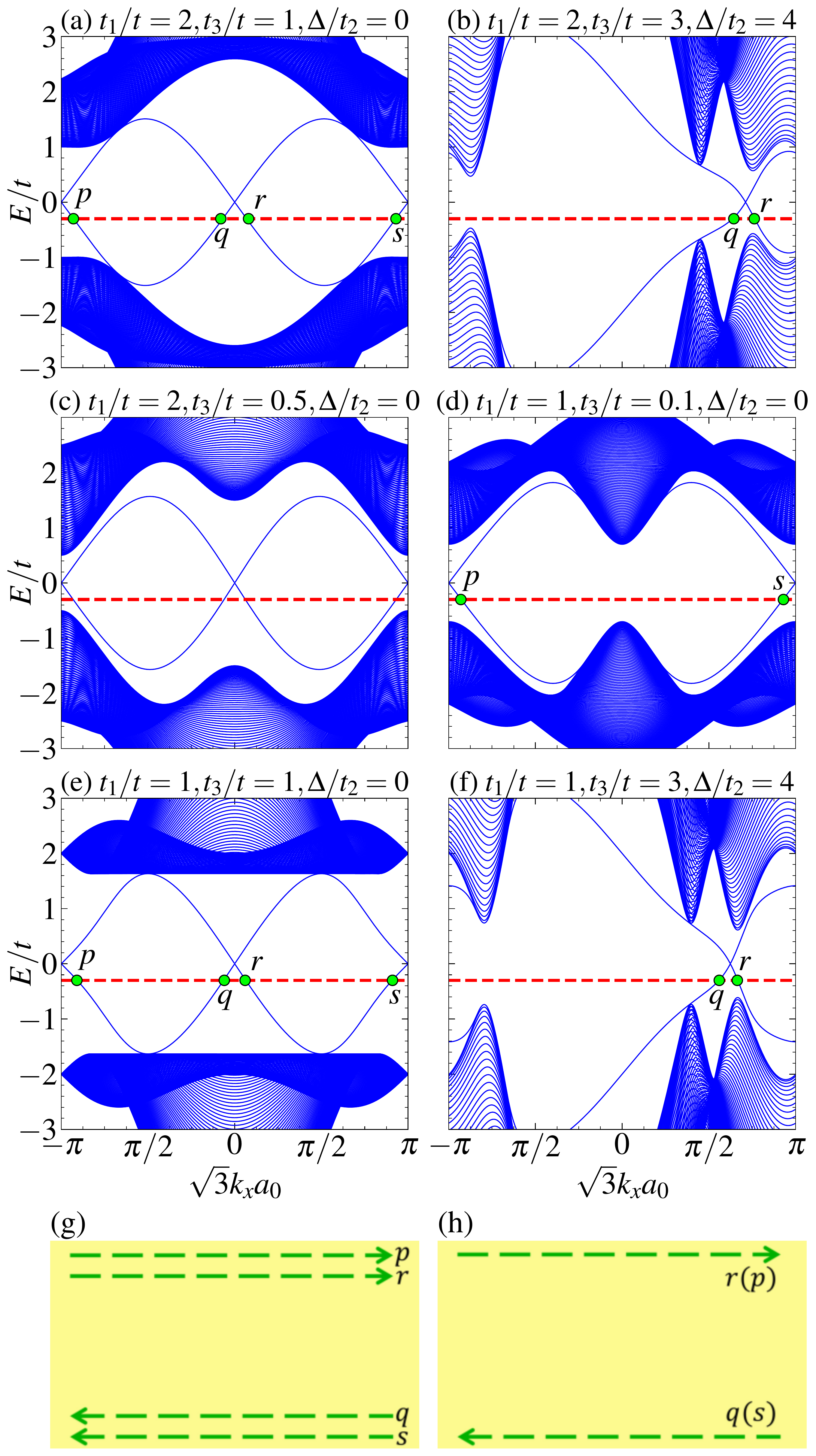}
		\begin{subfigure}{0\linewidth}
			\caption{}\label{fig:es_t1_2_t3_1_m_0}
		\end{subfigure}
		\begin{subfigure}{0\linewidth}
			\caption{}\label{fig:es_t1_2_t3_3_m_2}
		\end{subfigure}
		\begin{subfigure}{0\linewidth}
			\caption{}\label{fig:es_t1_2_t3_0.5_m_0}
		\end{subfigure}
		\begin{subfigure}{0\linewidth}
			\caption{}\label{fig:es_t1_1_t3_0.1_m_0}
		\end{subfigure}
		\begin{subfigure}{0\linewidth}
			\caption{}\label{fig:es_t1_1_t3_1_m_0}
		\end{subfigure}
		\begin{subfigure}{0\linewidth}
			\caption{}\label{fig:es_t1_1_t3_3_m_2}
		\end{subfigure}
		\begin{subfigure}{0\linewidth}
			\caption{}\label{fig:edge_current1}
		\end{subfigure}
		\begin{subfigure}{0\linewidth}
			\caption{}\label{fig:edge_current2}
		\end{subfigure}
		\caption{\raggedright The energy spectra of the ribbon as a function of the dimensionless momentum $k$ (here $k$ denotes $\sqrt{3}a_0 k_{x}$) for the semi-Dirac system ($t_1/t = 2$) are shown in (a) $t_3/t = 2$, $\Delta/t_2 = 0$ (b) $t_3/t = 3$, $\Delta/t_2 = 4$ and (c) $t_3/t = 0.5$, $\Delta/t_2 = 0$, while for the Dirac system ($t_1/t = 1$) it is shown in (d) $t_3/t = 0.1$, $\Delta = 0$, (e) $t_3/t = 1$, $\Delta = 0$ and (f) $t_3/t = 3$, $\Delta/t_2 = 4$. The green dots in each figure signify the intersection of the edge states with the Fermi energy $E_F$ (shown via the red dashed line). A schematic diagram  of a part of the ribbon is shown in (g) and (h). The arrows in (g) represent the edge currents corresponding to the points of figure (a) and (e), while in (b) the edge currents are shown corresponding to the points of figure (b), (d) and (f).}
		\label{fig:edge_states}
	\end{figure}

	\section{Edge states}\label{sec:edgestates}
	In order to understand whether the nature of the band gaps are topological or trivial, we look for the existence (or absence) of the edge states. To achieve this, we have considered the system to have semi-infinite ribbon geometry. Such a scenario breaks the periodicity along a particular direction, while the translational symmetry is preserved along the perpendicular direction. We take the semi-infinite ribbon \cite{nakada1996,castroneto,sticlet2012} to be finite along the $y$-direction, and infinite along the $x$-direction. We further label the sites along the $y$-direction as A$_1$, B$_1$, A$_2$, B$_2$, .... A$_N$, B$_N$ etc as shown in Fig. \ref{fig:edge_ribbon}. Since the translational invariance is preserved along the $x$-direction, we can fourier transform the operators along the $x$ direction only, that is, use $c_{x, y}^\dagger = \sum_{k} e^{ikx}c_{k, y}^\dagger$. This yields two sets of coupled eigenvalue equations for the wave functions which can be written as,

	\begin{equation}\label{eq:edge1}
		\begin{aligned}
			E_{k} a_{k, n} =&-\left[ t\left\{1+ e^{(-1)^n ik} \right\}b_{k, n} + t_1 b_{k, n-1} \right]\\ &-2t_2\left[ \cos(k + \phi)a_{k, n} + e^{(-1)^n\frac{ik}{2}}\times \right.\\ &\left. \cos\left( \frac{k}{2} - \phi\right) \{a_{k, n-1} + a_{k, n+1} \} \right]\\
			& +t_3 \left[b_{k, n+3} + 2 b_{k, n-1} \cos k\right]
		\end{aligned}
	\end{equation}
	\begin{equation}\label{eq:edge2}
		\begin{aligned}
			E_{k} b_{k, n} =&-\left[ t\left\{ 1+ e^{(-1)^{n+1} ik} \right\}a_{k, n} + t_1 a_{k, n+1} \right]\\ &-2t_2\left[ \cos(k - \phi)b_{k, n} + e^{(-1)^{n+1} \frac{ik}{2}}\times \right.\\ & \left. \cos\left( \frac{k}{2} + \phi\right)\{a_{k, n-1} + a_{k, n+1}\} \right]\\
			& +t_3 \left[a_{k, n-3} + 2 a_{k, n+1} \cos k\right]
		\end{aligned}
	\end{equation}
	\noindent where $n$ denotes the site index. $n$ assumes integer values in the range $[1:N]$ with $N$ being the total number of unit cells along the $y$-direction. In Eqs. \ref{eq:edge1} and \ref{eq:edge2}, $a_{k, n}$ and $b_{k, n}$ are the coefficients of the wave functions corresponding to the $n$-th A and B sublattices respectively. Here $k$ is the momentum along the periodic $x$-direction, which is rendered dimensionless by defining, $k = \sqrt{3} a_0 k_x$. The width $D$ of the ribbon along the $y$-direction is related to $N$ via $D(N) = a_0\left(\frac{3N}{2} -1\right)$. In our work, we have used $N = 256$ and hence the ribbon has a width of 383$a_0$. By solving Eqs. \ref{eq:edge1} and \ref{eq:edge2} one can get the bandstructure of the nanoribbon as shown in Fig. \ref{fig:edge_states} for a fixed value of the Haldane flux, namely $\phi = \pi/2$. As can be seen, one of the edge modes from the lower band crosses over to the upper band as a function of $k_x$, and another one crosses over in the opposite direction. These edge modes are responsible for a finite value of the Hall conductivity, provided the Fermi energy lies in the bulk gap.
	In Fig. \ref{fig:es_t1_2_t3_1_m_0}, we show the edge states for the semi-Dirac ($t_1 = 2t$) system corresponding to a particular value of the N3 hopping, for example, $t_3 = t$. The red dashed line represents the Fermi energy, $E_F$ and the points where the edge modes intersect the Fermi energy are shown by green dots. The edge currents corresponding to the points `$p$' and `$r$' flow along one edge of the ribbon and the edge currents corresponding to the points `$p$' and `$s$' travel along the other edge (see Fig. \ref{fig:edge_current1}). However their flow are in the opposite directions, since the velocity of the electron is proportional to $\partial E/\partial k$ which alters sign at $(q,s)$ compared to those at $(p, r)$. 
	
	Owing to the presence of a pair of edge states, there will be finite Hall conductivity with a plateau occurring at a value $2e^2/h$, with the factor `$2$' denoting the number of edge modes \cite{hatsugai1988}. This result is consistent with the Chern number phase diagram (see Fig. \ref{fig:pd_m_t3_t1_2}), where the Chern number is found to have a value -2 for $t_3 = t$ and $\Delta = 0$. In contrast, we get a single edge mode, along either edge of the ribbon for $t_3 = 3t$ and $\Delta = 4t_2$ as shown in Fig. \ref{fig:es_t1_1_t3_3_m_2}. In this case, we show the edge currents corresponding to the points `$q$' and `$r$' in Fig. \ref{fig:edge_current2}. This result is also consistent with the phase diagram (Fig. \ref{fig:pd_m_t3_t1_1}), where we find $C = 1$. For $t_2 = 0.5$ and $\Delta = 0$, the edge modes are shown in Fig. \ref{fig:es_t1_2_t3_0.5_m_0}. It is clearly visible that the edge modes have split from the bulk. Thus,  one can say that the edge modes do not contribute to the edge current and hence the system possesses zero Hall conductivity. 

	The edge states for the Dirac system ($t_1 = t$) are shown for comparison in Figs. \ref{fig:es_t1_1_t3_0.1_m_0}, \ref{fig:es_t1_1_t3_1_m_0} and \ref{fig:es_t1_1_t3_3_m_2}. The plots show the presence of single edge mode at the points `$p$' and `$s$' for $t_3 = 0.1t$ and $\Delta = 0$ (Fig. \ref{fig:es_t1_1_t3_0.1_m_0}), which propagates at two opposite edges of the ribbon corresponding to the points `$p$' and `$s$'. Such a situation yields a plateau in the Hall conductivity at $e^2/h$. This result is similar to that of the Haldane model. Now, if we increase the value of the N3 hopping, $t_3$ say, $t_3 = t$ (Fig. \ref{fig:es_t1_1_t3_1_m_0}), a pair of edge modes appear, and they propagate along two different edges of the ribbon, however these two pairs are counter propagating at the opposite edges (see Fig. \ref{fig:edge_current2}). In this case, the Hall plateau will be quantized in unit of $2e^2/h$. The pair of edge currents will be there as long as the Semenoff mass, $\Delta$ remains at a zero value. However, as we introduce a finite value of $\Delta$, there is possibility that there will be single edge mode  at each edge as depicted in Fig. \ref{fig:es_t1_1_t3_3_m_2}. Here we see that the edge modes are along either edge of the ribbon corresponding to the points `$q$' and `$r$' as shown in Fig. \ref{fig:edge_current2}. The number of edge currents along either edge of the ribbon is consistent with their values for the Chern numbers, namely, $C=1$ and $C = -2$ (see Fig. \ref{fig:pd_m_t3_t1_1}).

	\section{Anomalous Hall conductivity}\label{sec:hallcond}
	\begin{figure}[h]
		\begin{subfigure}[b]{0.23\textwidth}
			\includegraphics[width=\textwidth]{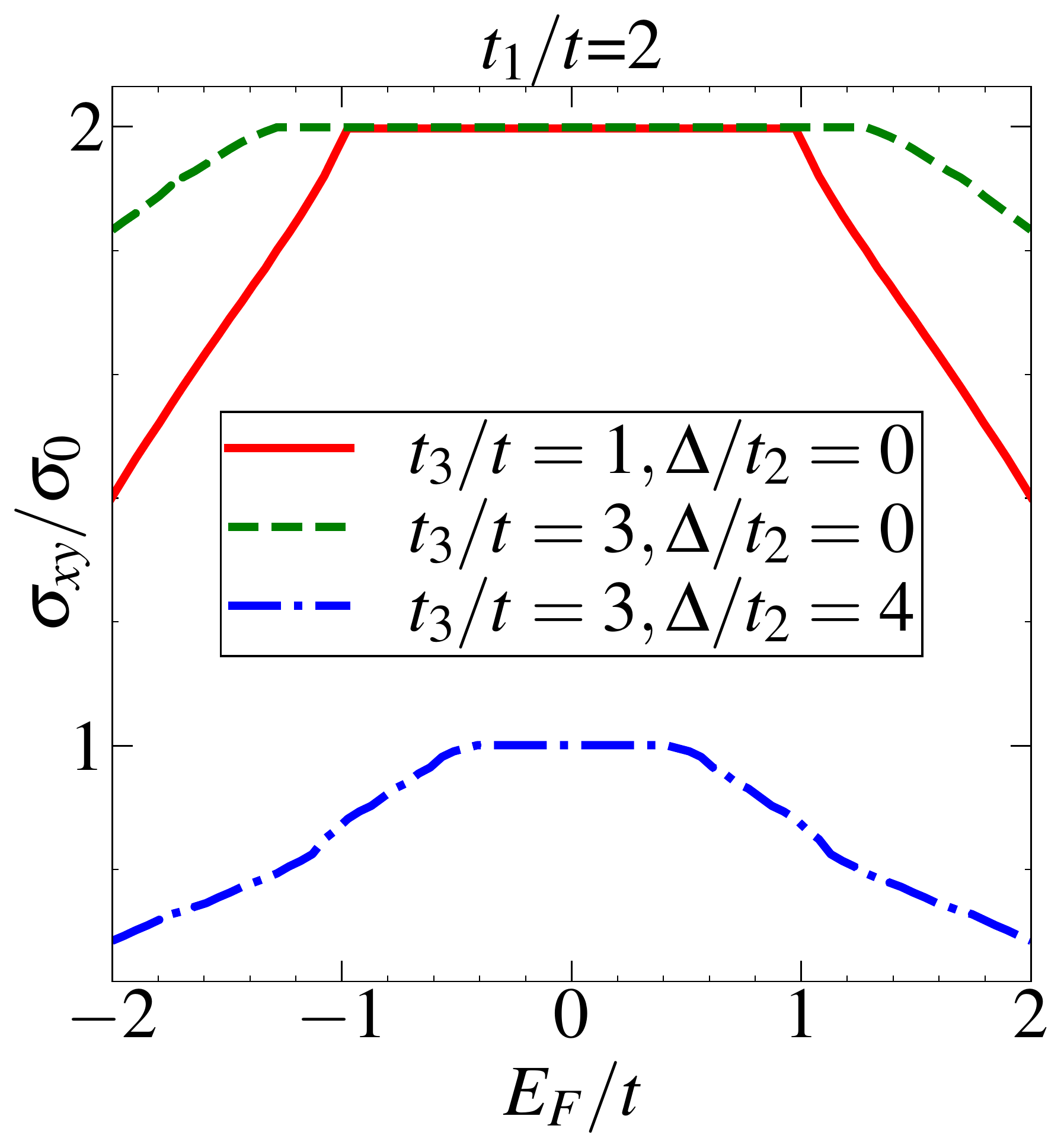}
			\subcaption{}\label{fig:hc_t1_2}
		\end{subfigure}
		\begin{subfigure}[b]{0.23\textwidth}
			\includegraphics[width=\textwidth]{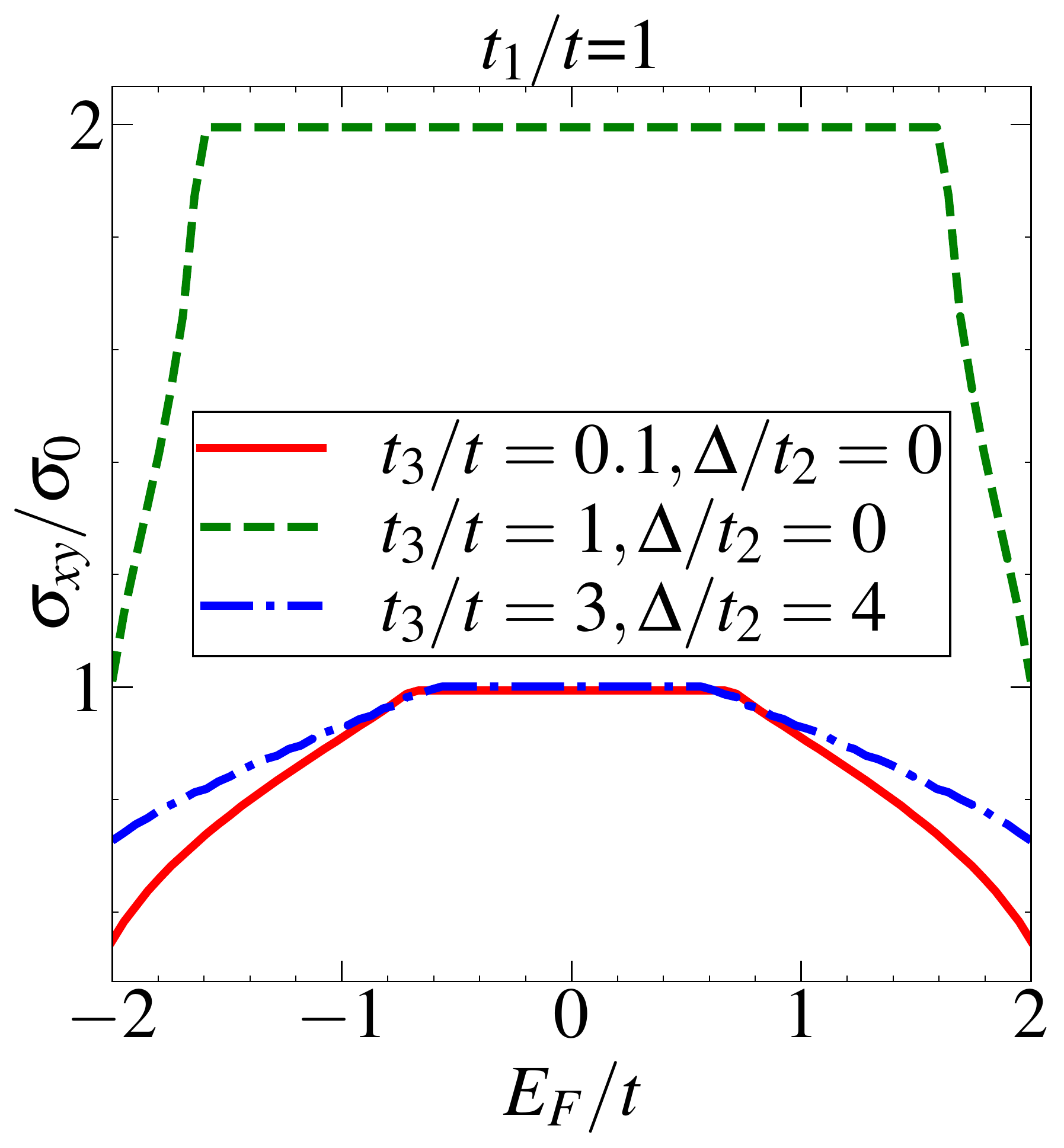}
			\subcaption{}\label{fig:hc_t1_1}
		\end{subfigure}
		\caption{\raggedright The variation of anomalous Hall conductivity, $\sigma_{xy}$ is shown as a function of the Fermi energy $E_F$ for (a) $t_1/t = 2$ and (b) $t_1/t = 1$. Here $\sigma_0 = e^2/h$ is the unit of the Hall conductivity. In this calculation we have fixed the N2 hopping $t_2$ at $0.5t$ and the Haldane flux $\phi$ at $\pi/2$.}\label{fig:hall_cond}
	\end{figure}
	The anomalous Hall conductivity requires a non-zero local Berry curvature. In order to calculate the Hall conductivity, we first obtain the Berry curvature of the system using Eq. \ref{eq:berry_curv} and then use the following formula \cite{hall1,hall2,kushsaha},
	
	\begin{equation}\label{eq:Hall_cond}
		\sigma_{xy} = \frac{\sigma_0}{2\pi} \sum_{\lambda} \int \frac{\mathrm{d}k_x \mathrm{d}k_y}{(2\pi)^2} f\left(E^\lambda_{k_x, k_y} \right) \Omega(k_x, k_y)
	\end{equation}
	 where $E^\lambda(k_x, k_y)$ denotes the electronic energies and $\lambda =$ $+1$ and $-1$ represent the upper and the lower bands respectively. $\sigma_0 (= e^2/h)$ sets the scale for $\sigma_{xy}$. $f(E) = \left[ 1+e^{(E-E_F)/K_BT} \right]^{-1}$ is the Fermi-Dirac distribution function with $E_F$ being the Fermi energy and $T$ is the absolute temperature. Using Eqs. \ref{eq:berry_curv} and \ref{eq:Hall_cond}, the Hall conductivity is calculated  numerically at zero temperature ($T = 0$) as a function of $E_F$ and are shown in Fig. \ref{fig:hc_t1_2} corresponding to the semi-Dirac system ($t_1 = 2t$). We see that as long the Fermi energy lies in the gapped region, the Hall conductivity shows a plateau quantized in unit of $2e^2/h$ for $t_3=t$ and $\Delta = 0$. Since the integral is performed over the occupied states for a given value of $E_F$, the Hall conductivity decreases as $E_F$ moves away from the gapped region, that is, towards the bulk. If we consider the Semenoff mass, $\Delta$ to be zero, we see that the plateaus occur at $2e^2/h$ as shown by the green and the red curves in Fig. \ref{fig:hc_t1_2}. However, in presence of a finite value of $\Delta$, there is a possibility of getting a plateau at $e^2/h$ occurring in $\sigma_{xy}$ as shown by the blue curve in Fig. \ref{fig:hc_t1_1}. These results are supported by the respective values for the Chern numbers. Thus, corresponding to $C=-2$, we get the Hall plateau quantized at a value $2e^2/h$, and for $C=1$, we get it is quantized at $e^2/h$. 
	 
	 To compare with the Dirac case, that is, for $t_1 = t$, we have shown the anomalous Hall conductivity in Fig. \ref{fig:hc_t1_1}. As can be seen, with a small N3 hopping (say, $t_3 = 0.1t$), the Hall plateau is quantized in unit of $e^2/h$ (the red curve in Fig. \ref{fig:hc_t1_1}). If we increase the value of the N3 hopping, the quantized Hall conductivity is seen at $2e^2/h$. Now, if we add the Semenoff mass term, the quantized Hall conductivity acquires a plateau at $e^2/h$. The value of $\Delta$ till which the $e^2/h$ plateau is retained depends on the value of $t_3$. In Fig. \ref{fig:hc_t1_1}, we show the Hall conductivity for a non-zero $\Delta$ by the blue curve corresponding to $t_3 = 3t$. The existence of the $e^2/h$ Hall plateau is noted for a certain range of $\Delta$, that is, $2.5t_2\lesssim\Delta\lesssim 3\sqrt{3}t_2$ corresponding to a fixed value of N3 hopping, namely, $t_3 = 3t$. Similar to the case of the semi-Dirac system, the quantized Hall conductivity of the Dirac system is fully consistent with the corresponding Chern number phase diagrams (see Fig. \ref{fig:pd_m_t3_t1_1}). 
	  
	\section{Conclusion}\label{sec:conclusion}
	We have shown that in the semi-Dirac system, adding a third neighbour hopping causes the zero modes to move inward into the BZ from its boundary. Addition of the Haldane term  creates spectral  gaps at those points. We have obtained two different phase diagrams, namely, in the parameter spaces defined by $\Delta$-$\phi$ and $\Delta$-$t_3$ by computing the Chern numbers. The $\Delta$-$t_3$ phase diagram is an addition to the existing literature. Further, the $\Delta$-$t_3$ phase diagram for the semi-Dirac case shows different scenario than the Dirac case in the following sense. There is always a trivial regime in between the two Chern insulating regimes ($C = -2$ and $C = 1$), which is absent for the Dirac case. The $\Delta$-$\phi$ phase diagram for the semi-Dirac case shows that one may have Chern insulating regions either with $|C| = 2$ and $|C| = 1$, or only $|C| = 2$, depending on the value of the N3 hopping, $t_3$. The computation of the edge states show additional crossing of the edge modes corresponding to $|C| = 2$. Finally, the anomalous Hall conductivities, for several values of $t_3$, demonstrate the existence of Hall plateaus quantized either at $ 2e^2/h$ or at $e^2/h$ depending on the values of their Chern number.


\begin{thebibliography}{8}
		\bibitem{klitzing}
		K. von Klitzing, \href{https://doi.org/10.1103/RevModPhys.58.519}{Rev. Mod. Phys. \textbf{58}, 519 (1986).}
		
		\bibitem{thouless1982}
		D. J. Thouless, M. Kohmoto, M. P. Nightingale, and M. den Nijs,
		\href{https://doi.org/10.1103/PhysRevLett.49.405}{Phys. Rev. Lett. \textbf{49}, 405 (1982).}
		
		\bibitem{thouless1983}
		D. J. Thouless, 
		\href{https://doi.org/10.1103/PhysRevB.27.6083}{Phys. Rev. B \textbf{27}, 6083 (1983).}
		
		\bibitem{avron1983}
		J. E. Avron, R. Seiler, and B. Simon, 
		\href{https://doi.org/10.1103/PhysRevLett.51.51}{Phys. Rev. Lett. \textbf{51}, 51 (1983).}
		
		\bibitem{kohomoto1985}
		M. Kohmoto,
		\href{https://doi.org/10.1016/0003-4916(85)90148-4}{Ann. Phys. (Berlin) \textbf{160}, 343 (1985).}
		
		\bibitem{Niu1985}
		Q. Niu, D. J. Thouless, and Yong-ShiWu, 
		\href{https://doi.org/10.1103/PhysRevB.31.3372}{Phys. Rev. B \textbf{31}, 3372 (1985).}
		
		\bibitem{prange1990}
		R. E. Prange and S. M. Girvin (eds.), \textit{The Quantum Hall Effect} (Springer-Verlag, Berlin, 1990).
		
		\bibitem{laughlin}
		R. B. Laughlin, \href{https://doi.org/10.1103/PhysRevB.23.5632}{Phys. Rev. B \textbf{23}, 5632(R) (1981).}
		
		\bibitem{trugman1983}
		S. A. Trugman, \href{https://doi.org/10.1103/PhysRevB.27.7539}{Phys. Rev. B \textbf{27}, 7539 (1983).}
		
		\bibitem{ilani2004}
		S. Ilani, J. Martin, E. Teitelbaum, J. H. Smet, D. Mahalu, V. Umansky, and A. Yacoby, 
		\href{https://doi.org/10.1038/nature02230}{Nature \textbf{427}, 328 (2004).}
		
		\bibitem{vasil1985}
		P. Vasilopoulos,
		\href{https://doi.org/10.1103/PhysRevB.32.771}{Phys. Rev. B \textbf{32}, 771 (1985).}
		
		\bibitem{tong}
		D. Tong, \href{https://arxiv.org/abs/1606.06687}{arXiv:1606.06687.}
		
		\bibitem{Haldane1988}
		F. D. M. Haldane,
		\href{https://doi.org/10.1103/PhysRevLett.61.2015}{Phys. Rev. Lett. {\bf 61}, 2015 (1988).}
		
		\bibitem{semenhoff}
		G. W. Semenoff,
		\href{https://doi.org/10.1103/PhysRevLett.53.2449}{Phys. Rev. Lett. {\bf 53}, 2449 (1984).}
		
		\bibitem{vanderbilt2006}
		T. Thonhauser and D. Vanderbilt,
		\href{https://doi.org/10.1103/PhysRevB.74.235111}{Phys. Rev. B {\bf 74}, 235111 (2006).}
		
		\bibitem{hasan2010}
		M. Z. Hasan and C. L. Kane, 
		\href{https://doi.org/10.1103/RevModPhys.82.3045}{Rev. Mod. Phys. \textbf{82}, 3045 (2010).}
		
		\bibitem{ando2013}
		Y. Ando, \href{https://doi.org/10.7566/JPSJ.82.102001}{J. Phys. Soc. Jpn. \textbf{82}, 102001 (2013).}
		
		\bibitem{qi2011}
		X.-L. Qi and S.-C. Zhang, \href{https://doi.org/10.1103/RevModPhys.83.1057}{Rev. Mod. Phys. \textbf{83}, 1057 (2011).}
		
		\bibitem{moore2010}
		J. E. Moore, \href{https://doi.org/10.1038/nature08916}{Nature (London) \textbf{464}, 194 (2010).}
		
		\bibitem{kim2017}
		H. S. Kim and H. Y. Kee,
		\href{https://doi.org/10.1038/s41535-017-0021-z}{npj Quant Mater \textbf{2}, 20 (2017)}
		
		\bibitem{Kapri2020}
		B. Dey, P. Kapri, O. Pal, and T. K. Ghosh,
		\href{https://doi.org/10.1103/PhysRevB.101.235406}{Phys. Rev. B {\bf 101}, 235406 (2020).}
		
		\bibitem{Dietl2008}
		P. Dietl, F. Pichon, and G. Montambaux, 
		\href{https://doi.org/10.1103/PhysRevLett.100.236405} {Phys. Rev. Lett. {\bf 100}, 236405    (2008).}
		
		\bibitem{Banerjee2009}
		S. Banerjee, R. R. P. Singh, V. Pardo, and W. E. Pickett,
		\href{https://doi.org/10.1103/PhysRevLett.103.016402} {Phys. Rev. Lett. {\bf 103}, 016402  (2009).}
		
		\bibitem{zieglar2017}
		K. Zieglar and A. Sinner, 
		\href{https://iopscience.iop.org/article/10.1209/0295-5075/119/27001}{EPL, {\bf 119}, 27001 (2017).}
		
		\bibitem{castro_2014}
		A. S. Rodin, A. Carvalho, and A. H. Castro Neto,
		\href{https://doi.org/10.1103/PhysRevLett.112.176801}{Phys. Rev. Lett. {\bf 112}, 176801 (2014).}
		
		\bibitem{guan_2014}
		J. Guan, Z. Zhu, and D. Tománek,
		\href{https://doi.org/10.1103/PhysRevLett.113.046804}{Phys. Rev. Lett. {\bf 113}, 046804 (2014).}
		
		\bibitem{katnelson_2015}
		A. N. Rudenko, Shengjun Yuan, and M. I. Katsnelson,
		\href{https://doi.org/10.1103/PhysRevB.92.085419}{Phys. Rev. B {\bf 92}, 085419 (2015).}
		
		\bibitem{katnelson_2016}
		C. Dutreix, E. A. Stepanov, and M. I. Katsnelson,
		\href{https://doi.org/10.1103/PhysRevB.93.241404}{Phys. Rev. B {\bf 92}, 241404(R) (2016).}
		
		\bibitem{pickett2009}
		V. Pardo and W.E. Pickett,
		\href{https://doi.org/10.1103/PhysRevLett.102.166803}{Phys. Rev. Lett. {\bf 102}, 166803 (2009).}
		
		\bibitem{pickett2010}
		V. Pardo, W.E. Pickett,
		\href{https://doi.org/10.1103/PhysRevB.81.035111}{Phys. Rev. B {\bf 81}, 035111 (2010).}
		
		\bibitem{Suzumura2013}
		Y. Suzumura, T. Morinari, and F. Pi\'echon,
		\href{https://doi.org/10.7566/JPSJ.82.023708}{ J. Phys. Soc. Jpn. {\bf 82}, 023708 (2013).}
		
		\bibitem{hasegawa2006}
		Y. Hasegawa, R. Konno, H. Nakano, and M. Kohmoto,
		\href{https://doi.org/10.1103/PhysRevB.74.033413}{Phys. Rev. B {\bf 74}, 033413 (2006).}
		
		\bibitem{zhang_2017}
		C. Zhong, Y. Chen, Y. Xie, Y.-Y. Sun, and S. Zhang,
		\href{https://doi.org/10.1039/C6CP08439G}{Phys. Chem. Chem. Phys. {\bf {19}}, 3820 (2017).}
		
		\bibitem{montambaux_2009}
		G. Montambaux, F. Pi\'echon, J.-N. Fuchs, and M. O. Goerbig,
		\href{https://doi.org/10.1103/PhysRevB.80.153412}{Phys. Rev. B {\bf 80}, 153412 (2009).}
		
		\bibitem{kim2015}
		J. Kim, S. S. Baik, S. H. Ryu, Y. Sohn, S. Park, B.-G. Park, J. Denlinger, Y. Yi, H. J. Choi, and K. S. Kim, 
		\href{https://doi.org/10.1126/science.aaa6486}{Science \textbf{349}, 723 (2015).}
		
		\bibitem{mondal2021}
		S. Mondal, P. Kapri, B. Dey, T. K. Ghosh and S. Basu,
		\href{https://doi.org/10.1088/1361-648x/abe798}{J. Phys.: Condens. Matter \textbf{33}, 225504, (2021)}
		
		\bibitem{sinha2020}
		P. Sinha, S. Murakami, and S. Basu
		\href{https://doi.org/10.1103/PhysRevB.102.085416}{Phys. Rev. B \textbf{102}, 085416 (2020).}
		
		\bibitem{sticlet2013}
		D. Sticlet and F. Piéchon,
		\href{https://doi.org/10.1103/PhysRevB.87.115402}{Phys. Rev. B \textbf{87}, 115402 (2013).}
		
		\bibitem{bena2011}
		C. Bena and L. Simon,
		\href{https://doi.org/10.1103/PhysRevB.83.115404}{Phys. Rev. B\textbf{ 83}, 115404 (2011).}
		
		\bibitem{thouless}
		D. J. Thouless, \textit{Topological Quantum Numbers in Nonrelativistic Physics} \href{https://books.google.co.in/books?id=BgbtCgAAQBAJ}{(World Scientific, Singapore, 1998).}
		
		\bibitem{avron1988}
		J. E. Avron, L. Sadun, J. Segert, and B. Simon,
		\href{https://doi.org/10.1103/PhysRevLett.61.1329}{Phys. Rev. Lett. \textbf{61}, 1329 (1988).}
		
		\bibitem{nakada1996}
		K. Nakada, M. Fujita, G. Dresselhaus, and M. S. Dresselhaus,
		\href{https://doi.org/10.1103/PhysRevB.54.17954}{Phys. Rev. B {\bf 54}, 17954 (1996).}
		
		\bibitem{castroneto}
		A. H. Castro Neto, F. Guinea, N. M. R. Peres, K. S. Novoselov, and A. K. Geim,
		\href{https://doi.org/10.1103/RevModPhys.81.109}{Rev. Mod. Phys. {\bf 81}, 109 (2009).}
		
		\bibitem{sticlet2012}
		D. Sticlet, F. Piéchon, J.-N. Fuchs, P. Kalugin, and P. Simon,
		\href{https://doi.org/10.1103/PhysRevB.85.165456}{Phys. Rev. B \textbf{85}, 165456 (2012).}
		
		\bibitem{hatsugai1988}
		Y. Hatsugai \href{https://doi.org/10.1103/PhysRevB.48.11851}{Phys. Rev. B \textbf{48}, 11851 (1988)}
		
		\bibitem{hall1}
		D. Xiao, M.-C. Chang, and Q. Niu,
		\href{https://doi.org/10.1103/RevModPhys.82.1959}{Rev. Mod. Phys. {\bf 82}, 1959 (2010).}
		
		\bibitem{hall2}
		D. Culcer, A.MacDonald, and Q. Niu,
		\href{https://doi.org/10.1103/PhysRevB.68.045327}{Phys. Rev. B {\bf 68}, 045327 (2003).}
		
		\bibitem{kushsaha}
		K. Saha,
		\href{https://doi.org/10.1103/PhysRevB.94.081103}{Phys. Rev. B {\bf 94}, 081103(R) (2016).}
		
		
	\end{thebibliography}
\end{document}